\documentclass[final,1p,times,twocolumn]{elsarticle}
\usepackage{graphicx}
\usepackage{caption}
\usepackage{xcolor}
\usepackage{textcomp}
\usepackage{cancel}
\usepackage[labelformat=parens]{subcaption}
\usepackage{array,multirow}
\usepackage{slashed}
\usepackage{tabularx}
\usepackage{hyperref}
\usepackage{amsmath}
\usepackage{amssymb}
\usepackage{textcomp}
\usepackage{array,multirow}
\usepackage{amssymb}
\usepackage{lineno,hyperref}
\usepackage{lineno}
    \nolinenumbers

\journal{Nuclear Physics B}

\begin{document}

\begin{frontmatter}

\title{Scalar Triplet Leptogenesis with a CP violating phase}


\author[mymainaddress]{Sreerupa Chongdar}
\ead{518PH1002@nitrkl.ac.in}

\author[mymainaddress]{Sasmita Mishra\corref{mycorrespondingauthor}}
\cortext[mycorrespondingauthor]{Corresponding author}
\ead{mishras@nitrkl.ac.in}

\address[mymainaddress]{Department of Physics and Astronomy, National Institute of Technology Rourkela, Sundargarh, Odisha, India, 769008}

\begin{abstract}
We study baryogenesis through leptogenesis via decay of triplet scalars embedded
in the Standard Model. We consider two triplets scenario where the vacuum expectation
value developed by one of the triplets is complex. The coupling of the triplet Higgs
with the Standard Model leptons and Higgs scalar allows two decay channels,
with branching ratios 
$B_L$ and $B_\phi$, respectively. It is known that the hierarchy
between the two branching ratios ($B_L \gg B_\phi$ or $B_\phi\gg B_L$), 
is sensitive to the generation of adequate CP violation and
efficiency of leptogenesis. Working in a hierarchical limit of branching 
ratios and requiring adequate CP violation, we find the mass of the 
lightest triplet can be as low as $1 \times 10^{10}$ GeV. In the 
temperature regime $\left[ 10^{9}, 10^{12} \right]$ GeV, flavor 
effects, especially two-flavor effects become important in the 
study of leptogenesis. In two-flavored regime we study flavor
effects in leptogenesis. It is observed that adequate baryon
asymmetry can not be obtained for purely flavored leptogenesis,
which corresponds to the hierarchical branching ratios, 
$B_L \gg B_\phi$. But the hierarchy in the other way 
$B_\phi \gg B_L$ still allows successful
baryogenesis through leptogenesis. The phase of the 
complex vacuum expectation value of the triplet scalar
is constrained by requiring non-zero CP violation. Considering
the mass scale of triplet scalars as TeV scale, we also study
 resonant leptogenesis in the light of TeV scale physics.
\end{abstract}


\end{frontmatter}
\nolinenumbers

\section{Introduction}
\label{sec:intro}
The origin of baryon asymmetry of the Universe (BAU) continues
to be one of the most important questions of fundamental physics. 
The observations from big bang nucleosynthesis and cosmic microwave
background (CMB) quantify the asymmetry 
\cite{WMAP:2003elm, WMAP:2008lyn,WMAP:2008ydk, 
Dunkley:2008ie,Planck:2015fie, Planck:2018vyg}, 
$\eta_B=\frac{(n_B-n_{\overline{B}})}{n_\gamma}$ as  
$4.7\times 10^{-10}\leq\eta_{B}\leq 6.5\times 10^{-10}$,
where $n_{B}$, $n_{\overline{B}}$, and $n_{\gamma}=
\frac{2\zeta(3)}{\pi^{2}}T^{3}$ denote the number densities 
of baryons, antibaryons and gamma photons, respectively.
The Sakharov conditions \cite{Sakharov:1967dj} provide necessary
ingredients for    dynamical generation of BAU viz; baryon number
(B) violation, C (charge conjugation) and
CP (charge conjugation and parity) violation and out of 
equilibrium dynamics in the processes involving baryons.
Although these ingredients are all present in the Standard 
Model (SM),  the CP violation is not adequate and for the 
departure from thermal equilibrium, the
electroweak phase transition is not strongly first order.
So one needs to go
beyond the SM to introduce new sources of CP violation and
provide a departure from thermal equilibrium. Apart from BAU,
there are other avenues where one needs
to contemplate beyond the SM. One such avenue is the results from
neutrino oscillation experiments that are the clearest evidence of
the existence of new physics. The experiments based on the neutrino
oscillation phenomenon, have established the fact that this abundant
particle in the Universe, has non-zero mass 
\cite{Fukuda:1998mi,Ahmad:2002jz,Ahmad:2002ka,Bahcall:2004mz}. 
The allowed ranges of neutrino oscillation parameters; 
three mixing angles, $\theta_{12}$, $\theta_{23}$, $\theta_{13}$, 
two mass squared differences $\Delta m^{2}_{21}$, $|\Delta m_{31}^{2}|$ 
($\Delta m^{2}_{ij} = m_i^2 -m_j^2$) and one Dirac CP-violating 
phase $\delta$ by latest neutrino oscillation experiments, 
for normal ordering of masses, are given in table-\ref{neop} 
\cite{Esteban:2020cvm}.
Also, the sum of neutrino masses is bounded below by the recent 
Planck bound $\sum m_i < 0.12$ eV \cite{Aghanim:2018eyx}. 
This is in contrast to the prediction of the SM, where neutrinos
are massless. Subsequently, one relies on theories beyond
the SM to circumvent the problem. There are attractive neutrino
mass models
where neutrino masses arise via seesaw mechanism by adding 
heavy states; right-handed neutrinos (type-I) 
\cite{Minkowski:1977sc,Yanagida:1979as,Mohapatra:1979ia}, 
scalar triplets (type-II) \cite{Magg:1980ut,Lazarides:1980nt,Mohapatra:1980yp,Cheng:1980qt}
 or fermion triplets (type-III) 
\cite{Foot:1988aq,Ma:1998dn} and many more variants. 
In models with seesaw mechanism, the interaction
of heavy states with the SM states originates neutrino 
mass along with lepton number ($L$) asymmetry
owing to the decays of the heavy states to the SM lepton(s). 
The out of equilibrium decays of the heavy states generate BAU 
through leptogenesis mechanism \cite{Fukugita:1986hr}. 
The lepton asymmetry is converted to baryon asymmetry by
$B-L$ conserving but $B+L$ violating sphaleron process.
In this framework, the Sakharov conditions are successfully met. Thus,
leptogenesis realizes a non-trivial link between two completely 
independent experimental observations: BAU and neutrino mixing
and hence mass. In the present
work, we study baryogenesis through leptogenesis in the type-II
seesaw model where the SM is
augmented with two scalar triplets, with one of the triplets
acquiring complex vacuum expectation value (vev).

 The type-II seesaw mechanism is special from other variants in two ways.
 First, the one-to-one relation between neutrino mass matrix and triplet
 Yukawa coupling makes it easy to detect the flavor structure of the 
 theory directly from the neutrino mass matrix, which can be reconstructed
 from the low energy data. Second, only one triplet scalar can generate three
 light neutrino masses. But with this setup, there is a drawback with leptogenesis. 
In the presence of complex couplings of the triplet with the SM Higgs 
and the leptons, the leptonic CP asymmetry is generated only at higher loops and is
highly suppressed. Therefore, new sources (addition of heavy right 
handed neutrino/ triplet scalar) of CP violation are required for
successful leptogenesis in the type-II framework. The addition of 
another triplet is supported by another ground. Although
substantial evidence has been collected from experiments in support of
neutrino oscillation, a lot remains to be known including absolute 
values of masses, mass generation mechanism, and nature of neutrinos: Dirac or Majorana.
In the absence of a clear theoretical understanding, various texture zeros
in the neutrino mass matrix have been investigated. The zero entries are often
backed by some underlying flavor symmetry in the lepton sector. 
Also, the reduced number of free parameters in the neutrino mass matrix enhances the
predictivity. In this context, two-zero textures have gathered considerable
attention.
Two zero texture \cite{Frampton:2002yf,Fritzsch:2011qv} in the 
lepton mass matrix arising from some underlying
flavor symmetry render useful studies. One such
two zero texture mass matrix represented as $B_2$, 
where $e-\mu$ and $\tau-\tau$ elements of neutrino mass 
matrix are zero. This texture naturally emerges by adding 
two triplet scalars in the SM and by imposing a horizontal 
symmetry, ${\mathcal{Z}_3}$ \cite{Grimus:2004az,Fritzsch:2011qv}. 
The authors in the references \cite{Meloni:2014yea,Singh:2019baq} study the
viability of two zero textures in the recent neutrino oscillation
experimental data and Planck \cite{Planck:2018vyg} data and 
show that the texture $B_2$ is consistent with data in the 
normal ordering of neutrino masses. 

In models where neutrino mass is generated through type-II seesaw mechanism,
the relevant part of the Lagrangian, related to generation of 
neutrino mass is the coupling of the triplet field, $\Delta$ to 
the SM lepton fields ($L$) through Yukawa type coupling, $Y_\Delta$ 
and that to the SM Higgs field ($\phi$) through scalar coupling, 
$\mu$. We call them $\Delta-L$ and $\Delta-\phi$ terms, respectively.
If the latter type of coupling is absent then the lepton number 
is violated
spontaneously by the vacuum expectation value (vev), $\omega$ of the neutral component of 
the scalar triplet. This leads to the existence of 
a massless scalar, Majoron, which would contribute to the 
invisible decay width of
the $Z$-boson. So this model is excluded by CERN LEP.
Also, in models of spontaneously broken CP symmetry, extended by triplet scalars, specific to our case, CP symmetry is imposed at the Lagrangian level,
making the Yukawa couplings real. In this case, spontaneous CP violation can be realized through the complex vevs developed by the triplet scalars of the model.  The light neutrino mass matrix being determined by the complex vevs of the triplets and real Yukawa matrices, CP violation  can be established at the low scale (such as neutrino oscillation experiments). The high energy CP violation required for leptogenesis mechanism can not be accommodated, in this case, as can be seen from equations (\ref{eq:LF}) and (\ref{eq:F}). However the model can be extended by singlet scalars to account for both low and high energy CP violation \cite{Branco:2012vs}. Some recent studies by adding scalar singlets to the Standard model that can give rise to spontaneous CP violation and light neutrino mass have been performed  in Ref. \cite{Barreiros:2020gxu} in the context of dark matter and in Ref. \cite{Barreiros:2022fpi}, in the context of
leptogenesis in vanilla type-I seesaw.
Inclusion of the $\Delta-\phi$ coupling term leads to 
the breaking of the lepton number explicitly and 
eliminates Majoron. Also, with this set up CP violation is achieved explicitly. The vev of the scalar triple, $\omega$ gives rise
to the mass matrix of the neutrinos, $m_{\nu_{ij}} = \sqrt{2}Y_{\Delta_{ij}} \omega$;
$\omega \approx \mu \upsilon^2/2M^2_{\Delta} $, where $\mu$ is the coupling
strength of $\Delta -\phi$ term and $\upsilon$ and $M_{\Delta}$
are the vev of the SM Higgs and mass of the triplet scalar, respectively. 
An upper limit on $\omega$ can be obtained from $\rho$-parameter constraint: ($\rho
=M_W^2/ M_Z^2 \cos^2 \theta_w$); $\omega/\upsilon \le 0.03$ or $\omega < 8$
GeV. 

A scalar triplet, unlike a fermion singlet, undergoes SM gauge interactions. 
The latter can keep the triplet close to thermal equilibrium at temperatures
$T \lesssim 10^{15}$ GeV. This can result in reduced efficiency of leptogenesis and hence seem
difficult to fulfill the out-of-equilibrium decay condition. However,
a more precise calculation, by solving the full set of
Boltzmann equations, shows that leptogenesis is efficient even at much lower temperature \cite{Hambye:2003ka, Hambye:2005tk, Hambye:2000ui}. The presence
of both $\Delta-L$ and $\Delta-\phi$ terms in the Lagrangian, allows
two different decays of triplet scalar and lepton number is violated only if both types
of decays co-exist. The interplay of the strength of the respective 
branching ratios, $B_L$ and $B_\phi$ plays an important role in 
successful baryogenesis through leptogenesis.
The amount of CP asymmetry, $\epsilon$, and efficiency, $\eta$ depend on $B_L$ and $B_\phi$ in
a different manner. The efficiency is maximal when $B_L \gg B_\phi $ or 
$B_\phi \gg B_L $, whereas this condition corresponds to suppressed CP violation
(as $\epsilon \propto\sqrt{B_L B_\phi}$). Solving a full set of 
Boltzmann equations, in a model independent way, it was
shown in Ref. \cite{Hambye:2005tk} that even for extremely hierarchical branching ratios, 
it is possible to achieve the required BAU.
 In Ref. \cite{Hambye:2005tk}, the authors also show,  
 the lower bound on the triplet mass and an upper bound on 
 the CP asymmetry assuming hierarchical  light neutrino 
 masses as $M_\Delta > 1.3\times 10^{11}$ GeV ($\tilde{m}_\Delta =
 0.05$ eV) and $M_\Delta > 2.8\times 10^{10}$ GeV ($\tilde{m}_\Delta = 0.001$ eV) ($\tilde{m}_\Delta^2 = {\rm Tr}(m_\Delta^\dagger m_\Delta$). Here, $m_\Delta$ is the triplet 
 contribution to the light neutrino masses, $m_\nu =m_\Delta +m_H$, 
 with $m_H$ as the contribution of the heavier particles to $m_\nu$.
In this work, we show, due to the interplay of 
complex phases of triplet vev and elements of the triplet Yukawa matrix, 
it is possible to saturate
the bound on the lightest triplet mass as $1 \times 10^{10}$ GeV while 
maintaining the hierarchical branching ratios and hence not negotiating 
on efficiency. Thus our study provides a possible realisation of the study in 
Ref. \cite{Hambye:2005tk}. The phase $\alpha$ is a free parameter,
nevertheless not all values of it, can account for an integrated framework of neutrino mass generation and leptogenesis. So, the study of leptogenesis constrains the value of $\alpha$. Also,
we study flavor effects in leptogenesis and find different cases of 
hierarchical branching
ratio (i) $B_\phi \gg B_L $ and (ii) $B_L \gg B_\phi$ give different results.
We find in the latter case which is
known as the purely flavored leptogenesis (PFL), the CP asymmetry parameter 
in our framework comes out
to be extremely small, and hence producing the required BAU may not be viable. 
Nevertheless, the other 
hierarchical limit in (i) stands viable. The phase of the complex vev of the 
triplet scalar can be constrained
by requiring non-zero CP violation and hence non-zero BAU.
For completeness, 
we also consider bringing down the mass scale of the triplet scalar
as low as TeV and study leptogenesis taking resonant effect and 
Sommerfeld correction into account.

Triplet scalars offer rich ground to study other
avenues like dark matter and inflation. 
The study of BAU, dark matter and neutrino mass involving triplet 
Higgs has been carried out in references
\cite{Parida:2020sng, Parida:2021wmd, Chakraborty:2019uxk,Mahapatra:2020dgk,AristizabalSierra:2012pv,Mishra:2019sye,Lu:2016dbc, Hati:2015hvq}. 
The possibility of explaining BAU, 
neutrino mass, and inflation simultaneously are studied 
in Ref. \cite{Barrie:2021mwi} by adding a triplet Higgs to 
the SM. Triplet leptogenesis in the Left-right symmetric
model is explored in references \cite{Hallgren:2007nq,Rink:2020uvt}.
The vev of the triplet scalar is usually considered to be real.
But this generalization does not hold always.
The complex nature of the scalar-scalar ($\Delta-\phi$ term) coupling also comes with the possibility of a complex vev of the triplet \cite{Ma:1998dx, Chaudhuri:2016rwo}. We explore this possibility here, considering
at least one of the triplet scalars develop a complex vev. The role of 
phase of the complex vev is important in LHC phenomenology as 
pointed out by the authors of  references \cite{Chaudhuri:2013xoa, Chaudhuri:2016rwo}. 
By considering the complex vev of one out of the two triplets,
the authors show that a spectacular signal in the form of Higgs 
plus the same-sign dilepton peak can be observed at LHC. 

The paper is organised as follows. In section (\ref{sec:model}) we discuss the light neutrino mass model with the SM
extended with two triplets and determine the unknown parameters from neutrino oscillation data. In section (\ref{sec:gen-disc}) we discuss the necessary ingredients for estimating the final baryon asymmetry
of the Universe. In section (\ref{sec:quant-bau}) we study the unflavored and flavored leptogenesis. The discussion on TeV scale leptogenesis is done in  section (\ref{sec:resonant}). And we outline our conclusions and future perspective in  section (\ref{sec:cncl}).
\section{Type-II Seesaw Mechanism with two triplet scalars}
\label{sec:model}
 In this section, we consider the SM  with one complex Higgs doublet, $\phi$ is extended  with two triplet scalars, $\Delta_1,\Delta_2$ with hypercharge ${\mathcal Y}=2$ \cite{Dev:2013ff,Chaudhuri:2013xoa, Chaudhuri:2016rwo}; $\Delta_i =(\delta_i^{++},\delta_i^+,\delta_i^0)$ with $i=1,2$.  The fields can be written in matrix form and are given by 
\begin{equation}
   \phi = 
\begin{pmatrix}
 \phi^+\\
 \phi^0
\end{pmatrix},\quad
\Delta_1=
   \begin{pmatrix}
   \frac{\delta^+_1}{\sqrt{2}} & \delta^{++}_1\\
    \delta^{0}_1 & -\frac{\delta^+_1}{\sqrt{2}}
  \end{pmatrix}, \quad
  \Delta_2=
  \begin{pmatrix}
   \frac{\delta^+_2}{\sqrt{2}} & \delta^{++}_2\\
    \delta^{0}_2 & -\frac{\delta^+_2}{\sqrt{2}}
  \end{pmatrix}.
  \end{equation}
  We can write the scalar potential in case of two triplet scalars, as 
\begin{eqnarray}
\nonumber
 V(\phi,\Delta_{1},\Delta_{2}) &=& a\phi^{\dag}\phi+ c (\phi^{\dag}\phi)^2+ \frac{b_{kl}}{2}{\rm Tr} (\Delta^{\dag}_k \Delta_l)+\frac{d_{kl}}{4}({\rm Tr}(\Delta^{\dag}_k \Delta_l))^2 +(e_{kl}-h_{kl})\phi^{\dag}\phi {\rm Tr}(\Delta^{\dag}_k \Delta_l) \\ \nonumber
 & + &  \frac{f_{kl}}{4} {\rm Tr}(\Delta^\dagger_{k} \Delta^\dagger_l){\rm Tr}(\Delta_k \Delta_l)+
 h_{kl}\phi^\dagger \Delta^\dagger_k \Delta_l \phi+ g {\rm Tr} (\Delta^{\dag}_1 \Delta_2){\rm Tr}(\Delta^{\dag}_2 \Delta_1) \\
 &+& g' {\rm Tr} (\Delta^{\dag}_1 \Delta_1){\rm Tr}(\Delta^{\dag}_2 \Delta_2)+(\mu_k\phi^{\dag} \Delta_k i\sigma_2\phi^{*}+ {\rm h.c.}),
 \label{vtt}
\end{eqnarray}
where $k,l=1,2$, and $\sigma_2$ is the second Pauli matrix. The vacuum expectation values  of the neutral components of the two triplet scalars and the Higgs doublet are given by 
\begin{equation}
 <\delta_1^0>=\omega_1,
  \quad
   <\delta_2^0>=\omega_2,
  \quad
  <\phi^0>=\frac{v}{\sqrt{2}},
\end{equation}
where $v=246$ GeV.  
In our study, we have taken the vevs; $\omega_1,\omega_2 \ll v$, keeping in mind the $\rho$-parameter constraint. There are two ways in which one can achieve this requirement theoretically by minimizing the scalar potential with respect to the vevs. One can assume $a, b_{kl} \sim v^2$; $c, d_{kl}, e_{hl}, f_{kl}, h_{kl}, g, g' \sim 1$; $|\mu_k| \ll v$ \cite{Grimus:1999fz,Chaudhuri:2016rwo} or in a more complete theory $b_{kl} \gg v^2 $ \cite{Grimus:1999fz, Ma:1998dx}. 

All parameters in the scalar potential are real except $\mu_k$, which is in general complex. The vev, $v$ can be chosen to be real by
performing a global $U(1)$ transformation but there is no second
global transformation that can make $\omega_{1(2)}$ real. For studying phenomenology related to CP violation 
and working with two triplet scalars, there lies an extra possibility of getting at least one CP-violating phase, which arises via a relative phase between the triplets \cite{Ma:1998dx}. For that purpose, any of the two scalar triplets is made to have a complex vev. Let it be the scalar triplet, $\Delta_1$. The coefficient of the corresponding trilinear term in the scalar potential is also complex, so one can write,
\begin{equation}
 \mu_1=|\mu_1|e^{i\beta}, \quad
 \omega_1=|\omega_1|e^{i\alpha}.
\end{equation}
For the sake of convenience, one can define the matrices,
\begin{equation}
 B = (b_{kl}), \quad E = (e_{kl}) \quad {\rm and} \quad H = (h_{kl}),
\end{equation}
and $\mu$ and $\omega$ can be written in terms of real (unprimed) and imaginary (primed) components in terms of vectors, 
\begin{equation} \mu=
  \begin{pmatrix}
   |\mu_1|\cos\beta \\
    \mu_2
  \end{pmatrix}, \quad
  \mu'=
   \begin{pmatrix}
   |\mu_1|\sin\beta \\
    0
  \end{pmatrix}, \quad
  \omega=
   \begin{pmatrix}
   |\omega_1|\cos\alpha \\
    \omega_2
  \end{pmatrix}, \quad
  \omega'=
   \begin{pmatrix}
   |\omega_1|\sin\alpha \\
    0
  \end{pmatrix}.
 \end{equation}
Now, minimising the potential in Eq.(\ref{vtt}), we get the following relations among the small vevs, 
\begin{equation}
 \omega=-v^2\left( B + \frac{1}{2} v^2 (E-H) \right)^{-1}\mu,
 \label{eq:small-vev}
\end{equation}
 and phases,
\begin{equation}
 \sin\beta=\frac{\left(b_{11}+ \frac{v^2}{2}(e_{11} -h_{11})\right) |\omega_1|\sin\alpha}{v^{2}|\mu_1|}
 \label{eq:beta}.
\end{equation}
 The relation in Eq.(\ref{eq:beta}) shows that the phases of $\mu_1$ and $\omega_1$ are related to each other. Also, the angle $\beta$ is absent, if the angle $\alpha$ takes the value $n\pi$, $n=0,1,2,3,\cdots$.
 
 The triplet scalars also couple to the SM leptons through Yukawa type interactions and the relevant part of the Lagrangian is given by,
\begin{equation} \mathcal{L}_Y=\sum_k \frac{1}{2}(Y_{\Delta})^k_{ij}L^T_iCi\sigma_2\Delta_k L_j+{\rm h.c.},
\end{equation}
where $Y_\Delta$ is Yukawa coupling matrix, $C$ is the charge conjugation matrix, $L_{i}$ denotes the left-handed lepton doublets, $i,j$ are the summation indices over the three neutrino flavors, and $k = 1,2$. The neutrino mass matrix can be written in terms of the Yukawa coupling constants and vev of the triplet's,
 \begin{equation}
  m_{\nu}= m^{(1)}_\nu + m^{(2)}_\nu =Y^{(1)}_{\Delta}|\omega_1|e^{i\alpha}+Y^{(2)}_{\Delta}\omega_2.
  \label{eq:mneu1}
 \end{equation}
%
 
The values of $\omega_1$, $\omega_2$ can be determined by the values of the other parameters in the scalar potential. But the Yukawa coupling constants are still undetermined. The Yukawa coupling matrix carries most of the unknown parameters, thus the theory lacks in making a definite prediction about fermion masses and flavor mixing.

 \subsection{Neutrino mass model with two scalar triplets}
 \label{sec:trip-model}
In this section, we make a quantitative analysis of the elements of the neutrino mass matrix given in Eq.(\ref{eq:mneu1}). We check the allowed range of elements of
the mass matrix in light of the recent experimental data related to neutrino mixing as a function of lightest neutrino mass considering normal ordering of masses. 

The general neutrino mass matrix for Majorana neutrinos is complex symmetric and can be 
 diagonalized as $m_\nu = U m_\nu^{\rm diag} U^T$, 
 where $m^{\rm diag}_{\nu}= {\rm Diagonal} ~(m_{1},m_{2}, m_{3} )$,
where $m_1, m_2$ and $m_3$ are the light neutrino mass eigenvalues.
In the basis where charged lepton mass matrix is real and diagonal, $U$ is the Pontecorvo-Maki-Nakagawa-Sakata (PMNS) matrix. The light neutrino mass matrix being complex symmetric, its reconstruction is possible 
by allowed experimental data. For example the $ee$ element of $m_\nu$ can be probed from neutrinoless double beta decay processes.
The decay widths of such processes are proportional to the square of the effective mass, $|m_{ee}| = |\sum_i m_i U_{ei}^2|$. For other elements of $m_\nu$, $m_{\alpha\beta}$ will govern 
 $\Delta L =2$ lepton number violating processes with charged leptons $\alpha$ and $\beta$ in the final state. 
 For example the rare decay $K^+ \rightarrow \pi^+ \mu^- \mu^-$, the branching ratio proportional to $|m_{\mu\mu}^2|$, where $m_{\mu\mu}$ is the $\mu\mu$ entry of $m_\nu$.
 There are also various studies on the aspects of such neurtino mass matrix where one admits texture zeros arising from symmetry principles. A study on the implications of vanishing matrix elements was conducted in Ref. \cite{Merle:2006du} by taking individual mass matrix elements as functions of the smallest
 neutrino mass. In the present paper we work with such a complex symmetric neutrino mass matrix.

 In our study, the light neutrino masses are generated through type-II seesaw mechanism, so there is a one-to-one correspondence between the triplet Yukawa couplings and light neutrino mass matrix (Eq.(\ref{eq:mneu1})). To realize this, we conjecture the triplet Yukawa matrices to take the form,
\begin{equation}
 Y^{(1)}_{\Delta} = 
 \begin{pmatrix}
 a e^{i\phi_{a}} & be^{i\phi_{b}} & ce^{i\phi_{c}}\\
  be^{i\phi_{b}} & de^{i\phi_{d}} & me^{i\phi_{m}}\\
  ce^{i\phi_{c}} & me^{i\phi_{m}} & ne^{i\phi_{n}}
 \end{pmatrix}, ~~
 Y^{(2)}_{\Delta} = 
 \begin{pmatrix}
  ae^{i\phi_{a'}} & be^{i\phi_{b'}} & ce^{i\phi_{c'}}\\
  be^{i\phi_{b'}} & de^{i\phi_{d'}} & me^{i\phi_{m'}}\\
  ce^{i\phi_{c'}} & me^{i\phi_{m'}} & ne^{i\phi_{n'}}
 \end{pmatrix},
 \label{eq:conjecture}
\end{equation}
such that the neutrino mass matrix in Eq.(\ref{eq:mneu1}) can be represented as, 
\begin{equation}
 m_\nu=
 \begin{pmatrix}
  Ae^{i\phi_{A}} & Be^{i\phi_{B}} & Ce^{i\phi_{C}}\\
  Be^{i\phi_{B}} & De^{i\phi_{D}} & Me^{i\phi_{M}}\\
  Ce^{i\phi_{C}} & Me^{i\phi_{M}} & Ne^{i\phi_{N}}
 \end{pmatrix},
 \label{eq:conj}
 \end{equation}
 where the explicit forms of the elements of $m_\nu$ are given in \ref{sec:elements}. 
In Eq.(\ref{eq:conjecture}), although the moduli of the corresponding elements of two Yukawa matrices are the same, their phases are taken to be different. In   \ref{sec:elements},
we show that the Yukawa matrices are different due to the associated phases, through detailed calculation. We have used
the benchmark values of different parameters that give
rise to successful BAU through leptogenesis.  A phenomenological  study, to an extent of pure phase matrices, has been summarized in Ref. \cite{Fritzsch:1999ee}.
 The phenomenology of such a
 complex symmetric matrix obtained in Eq.(\ref{eq:conj}) has been carried out in Ref. \cite{Adhikary:2013bma}.
%
The  PMNS matrix, $U$ is parameterized the  by three mixing angles $\theta_{ij}$,
one Dirac phase, $\delta$ and two Majorana phases $\alpha'$
and $\beta'$. In a standard parametrization it is given as, 
\begin{equation}
 U =
 \begin{pmatrix}
  c_{12}c_{13} & s_{12}c_{13} & s_{13}e^{-\delta}\\
  -s_{12}c_{23}-c_{12}s_{23}s_{13}e^{i\delta} & c_{12}c_{23}-s_{12}s_{23}s_{13}e^{i\delta} & s_{23}c_{13}\\
  s_{12}s_{23}-c_{12}c_{23}s_{13}e^{i\delta} & -c_{12}s_{23}-s_{12}c_{23}s_{13}e^{i\delta} & c_{23}c_{13}
 \end{pmatrix}
 {\rm diag} (1,e^{i\alpha'},e^{i(\beta'+\delta)}),
\end{equation}
where $s_{ij} (c_{ij}) \equiv \sin\theta_{ij} (\cos\theta_{ij})$.
Now the elements of the neutrino mass matrix  can be expressed as:
\begin{equation}
  m_{ee}= Ae^{i\phi_{A}}= c^{2}_{13}(m_{1}c^{2}_{12}+m_{2}e^{2i\alpha'}s^{2}_{12})+m_{3}e^{2i\beta'}s^{2}_{13},
  \label{m1}
 \end{equation}
 \begin{equation}
  m_{e\mu}= Be^{i\phi_{B}}=
c_{13}((e^{2i\alpha'}m_{2}-m_{1})s_{12}c_{12}c_{23}+e^{i\delta}(e^{2i\beta'}m_{3}-e^{2i\alpha'}m_{2}s^{2}_{12}-m_{1}c^{2}_{12})s_{23}s_{13}),
  \label{m2}
 \end{equation}
 \begin{equation}
  m_{e\tau}= Ce^{i\phi_{C}}=
  c_{13}((m_{1}-e^{2i\alpha'}m_{2})s_{12}c_{12}s_{23}+e^{i\delta}(e^{2i\beta'}m_{3}-e^{2i\alpha'}m_{2}s^{2}_{12}-m_{1}c^{2}_{12})c_{23}s_{13}),
  \label{m3}
 \end{equation}
 \begin{equation}
 m_{\mu\mu}= De^{i\phi_{D}}=
 m_{1}(s_{12}c_{23}+c_{12}s_{23}s_{13}e^{i\delta})^{2}+m_{2}(c_{12}c_{23}-s_{12}s_{23}s_{13}e^{i\delta})^{2}e^{2i\alpha'}+m_{3}s^{2}_{23}c^{2}_{13}e^{2i(\beta'+\delta)},
 \label{m5}
 \end{equation}
 \begin{eqnarray}
  m_{\mu\tau} &=&  Me^{i\phi_{M}}=\frac{1}{2}\sin 2\theta_{12}\cos 2\theta_{23}s_{13}e^{i\delta}(m_{1}-m_{2}e^{2i\alpha'}){}\\ & & {}-\frac{1}{2}\sin 2\theta_{23}(m_{2}c^{2}_{12}e^{2i\alpha'}+m_{1}s^{2}_{12}-e^{2i\delta}(m_{3}c^{2}_{13}e^{2i\beta'}+s^{2}_{13}(m_{1}c^{2}_{12}+m_{2}e^{2i\alpha'}s^{2}_{12}))),\nonumber
  \label{m4}
 \end{eqnarray}
 \begin{equation}
  m_{\tau\tau}= Ne^{i\phi_{N}} =m_{1}(c_{12}c_{23}s_{13}e^{i\delta}-s_{12}s_{23})^{2}+m_{2}(c_{12}s_{23}+s_{12}c_{23}s_{13}e^{i\delta})^{2}e^{2i\alpha'}+m_{3}c^{2}_{23}c^{2}_{13}e^{2i(\beta'+\delta)}.
  \label{m6}
 \end{equation}

 \begin{table}
 \centering 
 \begin{tabular}{ |p{3cm}||p{3cm}||p{3cm}|  }
 \hline
 \multicolumn{3}{|c|}{NuFIT 5.0 (2020)} \\
 \hline
 Neutrino Oscillation parameters& Normal ordering 1$\sigma$ &Normal ordering 3$\sigma$\\
 \hline
 \hline
 $\theta_{12}/^{\circ}$   & $33.44^{+0.77}_{-0.74}$    &$31.27\rightarrow35.86$\\
 $\sin^{2}\theta_{12}$   & $0.304^{+0.012}_{-0.012}$    &$0.269\rightarrow0.343$\\
 $\theta_{23}/^{\circ}$   & $49.2^{+0.9}_{-1.2}$    &$40.1\rightarrow51.7$\\
 $\sin^{2}\theta_{23}$&   $0.573^{+0.016}_{-0.020}$  & $0.415\rightarrow0.616$   \\
 $\theta_{13}/^{\circ}$   & $8.57^{+0.12}_{-0.12}$    &$8.20\rightarrow8.93$\\
 $\sin^{2}\theta_{13}$ &$0.02219^{+0.00062}_{-0.00063}$ & $0.02032 \rightarrow 0.02410$\\
 $\delta/^{\circ}$    &$197^{+27}_{-24}$ & $120\rightarrow369$\\
 $\frac{\Delta m^{2}_{21}}{10^{-5}{\rm eV}^{2}}$&   $7.42^{+0.21}_{-0.20}$  & $6.82\rightarrow8.04$\\
 $\frac{\Delta m^{2}_{3l}}{10^{-3}{\rm eV}^{2}}$& $+2.517^{+0.026}_{-0.028}$  & $+2.435\rightarrow+2.598$   \\
 \hline
\end{tabular}
\caption{Allowed ranges of neutrino oscillation parameters for normal  mass ordering \cite{Esteban:2020cvm}.}
\label{neop}
\end{table}
 
From Eq.s (\ref{m1} - \ref{m6})
 we can see, $A = |m_{ee}|, B=|m_{e\mu}|, C = |m_{e\tau}|, D = |m_{\mu\mu}|, M = |m_{\mu\tau}|$ and $N = |m_{\tau\tau}|$. Through this one can determine the values of $a, b, c, d, m$ and $n$ up to the phase factors. To determine the values of the parameters ($A, B, C, D, M, N$), we take the allowed ranges of neutrino oscillation parameters (three mixing angles, $\theta_{12}$, $\theta_{23}$, $\theta_{13}$, two mass squared differences $\Delta m^{2}_{21}$, $\Delta m^{2}_{3l}$ and one Dirac CP-violating phase $\delta$) by latest neutrino oscillation experiments given in table-\ref{neop}. In our study, we assume normal ordering
 of mass eigen values. The allowed values of the parameters
 as a function of the lightest neutrino mass $m_1$, consistent with experimental data, are shown in Fig.(\ref{fig:mdnplots}). A similar analysis on elements of neutrino mass matrix is done in Ref. \cite{Merle:2006du}. In section (\ref{sec:quant-bau}), we calculate CP asymmetry using these values of the parameters that naturally arise in the formulation.
 \section{Baryogenesis through Leptogenesis}
 \label{sec:gen-disc}
In this section, we summarise the essential points of baryogenesis through leptogenesis from decay of triplet scalars, $\Delta_\alpha$, ($\alpha =1,2$) on a general ground. We lay down the important aspects of both flavored and unflavored leptogenesis, that go in to final estimation of baryon asymmetry. 
\subsection{CP asymmetry parameter}
 \label{sec:CPasym-unflav}  
\begin{figure}[h]
  \centering
 \begin{subfigure}{0.9\linewidth}
   \includegraphics[width=\linewidth]{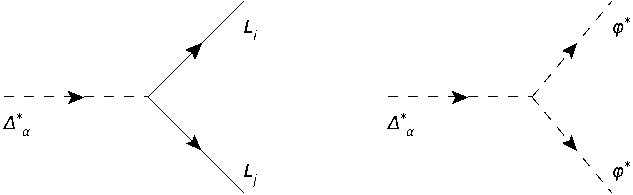}
   \caption{Tree level decay diagram: Triplet decaying into leptons and SM Higgs, respectively.}
   \label{fd1}
   \end{subfigure}
   \begin{subfigure}{0.9\linewidth}
   \includegraphics[width=\linewidth]{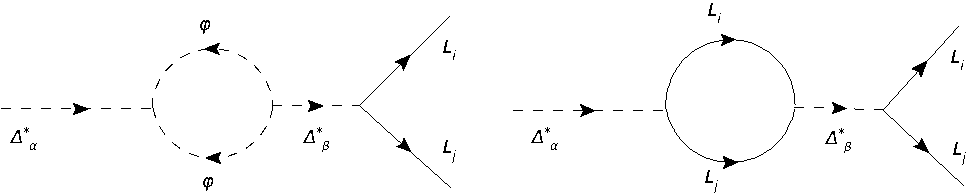}
   \caption{One-loop decay diagram: Triplet decaying into leptons through Higgs loop and lepton loop, respectively}
   \label{fd2}
   \end{subfigure}
   \caption{Feynman diagram for triplet Higgs scalar decaying into leptons and SM Higgs}
   \end{figure}

The gauge invariance of $SU(2)_L \times U(1)_{\mathcal{Y}}$ at an energy scale above electroweak symmetry breaking, ensures that we can pick up
any one of the three components of the triplet under consideration and the results are valid for the other two. By
choosing $\delta_\alpha^{++}$, the possible decay channels are,
\[
    \delta_\alpha^{++} \rightarrow 
\begin{cases}
     L_i^+\, L_j^+ & (L = -2)\\
    \phi^+ \, \phi^+   & (L = 0). 
\end{cases}
\]
The simultaneous existence of the above two final states
indicates that the lepton number is violated by two units ($\Delta L =2$). However, the lepton asymmetry generated by the decay of  $\delta_\alpha^{++}$ would be neutralized by the decays of 
$\delta_\alpha^{--}$ unless CP conservation is also violated and the decays
are out of thermal equilibrium in the early Universe. 
Before electroweak symmetry breaking the mixing between 
$\delta_1^0$ and $\bar{\delta}_2^0$ is strictly forbidden and
without loss of generality we can take the tree-level 
mass matrix, $b_{kl}$ to be real and diagonal. 
Hence, at this level, CP is conserved. However, CP violation can still take place at one loop level due to
the interference between tree and one-loop diagrams \cite{Ma:1998dx}.
In the presence of complex couplings of the triplets with the SM Higgs doublet and the leptons, a non-vanishing leptonic asymmetry is created for each component of triplet,
 \begin{equation}
  \epsilon^{L_{i}}_{\Delta_\alpha} = \Delta L\times 
  \frac{\sum_j \Gamma(\bar{\Delta}_\alpha \rightarrow L_i L_j) - \Gamma(\Delta_a\rightarrow \bar{L}_i \bar{L}_j)}{\Gamma_{\Delta_\alpha}+\Gamma_{\bar{\Delta}_\alpha}},
 \end{equation}
 where the decay width of the triplet scalar to the SM leptons is given by
 \begin{equation}
 \Gamma(\Delta_\alpha \rightarrow \bar{L}_i \bar{L}_j) =
 \frac{M_{\Delta_\alpha}}{8\pi} |Y^{(\alpha)}_{\Delta_{ij}}|^2
 \left[1+|Q-1|(1-\delta_{ij}) \right],
  \end{equation}
  where the electric charges of different components of 
  triplet, $\Delta_\alpha$ with masses $M_{\Delta_\alpha}$
  ($\alpha =1,2$) are represented as $Q$. Also,
  the branching ratio of triplet to the SM Higgs is given by,
\begin{equation}
 \Gamma(\Delta_\alpha \rightarrow \phi\phi) = \frac{|\mu_\alpha|^2}
 {8 \pi M_{\Delta_\alpha}}.
\end{equation}
 The branching ratios of the decay of the triplets to the leptons and Higgs are given as,
 \begin{equation}
  B^\alpha_L = \sum_{i = e, \mu, \tau} B^\alpha_{L_i}
  =\sum_{i,j= e,\mu, \tau} B^\alpha_{L_{ij}} = 
  \sum_{i,j= 1, 2, 3} \frac{M_{\Delta_\alpha}}{8\pi \Gamma_{\Delta_\alpha}} |Y^{(\alpha)}_{\Delta_{ij}}|^2, ~~ B^\alpha_\phi
  = \frac{|\mu_{\alpha}|^2}{8 \pi M_{\Delta_\alpha}
  \Gamma_{\Delta_\alpha}},
  \label{eq:br-ratio}
 \end{equation}
 where $\Gamma_{\Delta_\alpha}$ is the total decay width of the triplet,
\begin{equation}
  \Gamma_{\Delta_\alpha
  } = \frac{M_{\Delta_\alpha}}{8\pi}
  \left( \sum_{i,j}|Y^{(\alpha)}_{\Delta_{ij}}|^2 + 
  \frac{|\mu_{\alpha}|^2}{M^2_{\Delta_\alpha}}\right) \end{equation}
 The interference of the tree level (Fig. (\ref{fd1})) and one loop decay amplitudes gives rise to CP asymmetries. There can be two types of one-loop decays: one is mediated by the SM Higgs and the other is mediated by the leptons. They are shown in the Fig. (\ref{fd2}).
 In the case of the Higgs loop, it accounts for both total lepton number violation and lepton flavor violation. On the other hand, the one-loop mediated by leptons only causes lepton flavor violation. Based on these two contributions, the total CP asymmetry can be expressed as,
 \begin{equation}
  \epsilon^{L_{i}}_{\Delta_\alpha}=\epsilon^{{L_i} (\cancel{L},\cancel{F})}_{\Delta_\alpha}+\epsilon^{L_i (\cancel{F})}_{\Delta_\alpha},
  \label{eq:CP-tot}
 \end{equation}
where $\epsilon^{{L_i} (\cancel{L},\cancel{F})}_{\Delta_\alpha}$  and $\epsilon^{L_i( \cancel{F})}_{\Delta_\alpha}$ contributions come from the interference of tree-level diagrams with one loop diagrams of the Higgs loop and  the lepton loop, respectively. The corresponding contributions are given by,
\begin{equation}
 \epsilon^{L_{i}(\cancel{L},\cancel{F})}_{\Delta_\alpha}=\frac{1}{2\pi}\sum_{\beta \neq \alpha }  \frac{{\rm Im}[(Y^{(\alpha)\dagger}_{\Delta}Y^{(\beta)}_{\Delta})_{ii}\mu^*_\alpha
 \mu_\beta]}{M^{2}_{\Delta_{\alpha}}{\rm Tr}[Y^{(\alpha)\dagger}_\Delta Y^{(\alpha)\dagger}_\Delta]+|\mu_\alpha|^{2}}g\left(x \right),
 \label{eq:LF}
\end{equation}
\begin{equation}
 \epsilon^{L_{i}(\cancel{F})}_{\Delta_\alpha}=\frac{1}{2\pi}\sum_{\beta \neq \alpha } M^{2}_{\Delta_{\alpha}}\frac{{\rm Im}[(Y_\Delta^{(\alpha)\dagger}Y^{(\beta)}_{\Delta})_{ii}Tr[Y^{(\alpha)}_\Delta Y^{(\beta)\dagger}_{\Delta}]]}{M^{2}_{\Delta_{\alpha}}{\rm Tr}[Y^{(\alpha)}_\Delta Y^{(\alpha) \dagger}_\Delta]+|\mu_{\alpha}|^{2}}g\left(x \right),
 \label{eq:F}
\end{equation}
where 
\begin{equation*}
x=\frac{M^{2}_{\Delta_{\alpha}}}{M^{2}_{\Delta_{\beta}}},\quad g(x)=\frac{x(1-x)}{(1-x)^{2}+xy}, \quad
 y=\left(\frac{\Gamma^{tot}_{\Delta_{\beta}}}{M_{\Delta_{\beta}}}\right)^{2}.
\end{equation*}
The equations (\ref{eq:LF}) and (\ref{eq:F}) render interesting consequences specific to triplet leptogenesis. In this case the total lepton asymmetry can be written as
\begin{equation}
 \epsilon_{\Delta_\alpha} = \sum_{i = e,\mu, \tau} \epsilon^{L_i}_{\Delta_\alpha} = \sum_{i = e,\mu,\tau} \epsilon_{\Delta_\alpha}^{L_i (\cancel{L},\cancel{F})},
 \label{eq:CP-tot-higgs}
 \end{equation}
 since the other piece (Eq.(\ref{eq:F})) satisfies total lepton number conservation constraint,
 \begin{equation}
  \sum_i \epsilon_{\Delta_\alpha}^{L_i(\cancel{F})} = 0.
 \end{equation}
 It can also be seen that for the condition,
 \begin{equation}
  \mu^*_\alpha \mu_\beta \ll M^2_{\Delta_\alpha} {\rm Tr}\left[   Y^{(\alpha)}_\Delta Y^{(\beta)\dagger}_{\Delta} \right],
 \end{equation}
the entire CP asymmetry gets contribution mainly from the interference of tree level and one loop decay mediated by the leptons. This can be guaranteed by choosing the branching ratios to be $ B^\alpha_{L} \gg B^\alpha_{\phi}$. 
Hence the leptogenesis evolving from this CP asymmetry is called to be purely flavored leptogenesis. 

Finally for unflavored case the total CP asymmetry can be obtained by summing over all flavors as given in 
Eq.(\ref{eq:CP-tot-higgs}) and for hierarchical masses of the triplet scalars it is given by,
\begin{equation}
 \epsilon_ {\Delta_\alpha}=\frac{2 M_{\Delta_{\alpha}}(B^\alpha_{L}B^\alpha_{\phi})^{\frac{1}{2}}}{4\pi v^{2}}\times\frac{ {\rm Im}[(m^{(\alpha)}_{\nu})_{ij}(m^{(\beta)*}_{\nu})_{ij}]}{[{\rm Tr} (m^{(\alpha)\dagger}_{\nu}m^{(\alpha)}_{\nu})]^{\frac{1}{2}}}.
\label{eps}
\end{equation}
The dependence of the branching ratios, $B^\alpha_L$ and $B^\alpha_\phi$ on the
CP asymmetry parameter has interesting consequences, which we explore in the upcoming sections.
\subsection{Boltzmann equations}
\label{sec:BE-discuss}
In this section, we summarise and put together all the Boltzmann equations to compute the final lepton asymmetry
and hence, the baryon asymmetry. We assume that there was no lepton
asymmetry when the temperature of the Universe was much higher than the masses of triplets.
As the temperature reached the mass scale of one of the heavier triplets, lepton asymmetry was created due to the decay of the heavier triplet, $\Delta_1$ in our case.
The lepton asymmetry produced in the $\Delta_1$ decays would be washed out by the lepton number non-conserving interactions of $\Delta_2$. Thus, the effects of lepton number violating processes due to $\Delta_1$, can be safely neglected. Only the lepton number asymmetry arising from the interactions of $\Delta_2$ would evolve with time. From now onward we
represent $\Delta_2$ as $\Delta$ with mass $M_\Delta$ for our study and the corresponding scalar coupling $\mu_2$ as $\mu$. Also, the branching ratios $B^\alpha_L (B^\alpha_{L_i})$ and $B^\alpha_\phi$ are represented as $B_L (B_{L_i})$ and $B_\phi$, respectively. The CP asymmetry parameter, $\epsilon^{L_i}_{\Delta_\alpha}(\epsilon_{\Delta_\alpha})$ is represented as $\epsilon^{L_i}(\epsilon)$.
We intend to bring down the mass scale of triplet, $M_\Delta$ and hence the leptogenesis scale as low as possible. So the study requires a detailed formulation of Boltzmann equations.

At temperature, $T \gtrsim 10^{15}$ GeV all the SM reactions are slow and not in thermal equilibrium and 
only the interactions related to triplet are relevant. Only top quark ($t$) related Yukawa interactions enter thermal equilibrium for $T \subset [10^{12}, 10^{15}]$ GeV. The Yukawa
interactions related to bottom, charm, and tau, electroweak sphalerons, and QCD instantons
enter thermal equilibrium for $T\subset [10^9, 10^{12}]$ GeV. In this temperature regime, the processes induced by $\tau-$Yukawa are in thermal equilibrium and it breaks the coherence between $\tau$-lepton and the other two leptons (e and $\mu$). Also, electroweak sphalerons are
in thermal equilibrium in this temperature regime. Baryon number is no longer conserved while individual $B/3 - L_i$ is
conserved. So the evolution $B-L$ asymmetry should be done by following the evolution
of the flavored charge symmetries $B/3 - L_i$, where $i = a,\tau$; $a$ being the coherent
superposition of $e$ and $\mu$ lepton flavors. The strange and muon Yukawa interactions
resume thermal equilibrium in the range $T\subset [10^5, 10^9]$ GeV. In this regime, the
lepton doublets completely lose their quantum coherence. Finally, all the SM interactions are
in thermal equilibrium for $T\lesssim 10^5$ GeV.
 
The flavor decoherence requires careful quantum treatment. However, a simplified treatment can be given considering two relevant processes: (i) the SM lepton Yukawa interactions ($L-\phi$) and (ii) lepton related triplet inverse decays, ($L-\Delta$). Consider the situation where at a certain time the $L-\phi$ interaction rate becomes faster than the Hubble rate but $L-\Delta$ processes are much faster than (i). In this case, the coherent superposition of leptons produced from the decay of $\Delta$ will inverse decay before it has
the time to undergo $L-\phi$ interaction. In this case flavor decoherence has to wait until
the reduction in the temperature of the hit bath when the $L-\Delta$ interaction in (ii) is Boltzmann
suppressed.
 
Below $T = 10^{12}$ GeV the Yukawa induced processes like 
$q_3 L_{\tau}\rightarrow t_R \tau_R$ are in equilibrium, so the coherence between $\tau$-leptons
and other two flavors is broken.  So the coherence of $L_a = (Y_{1e} L_e + Y_{1\mu}L_\mu)/\sqrt{|Y_{1e}|^2 + |Y_{1\mu}|^2}$ is preserved. In this temperature regime $T\subset [10^9, 10^{12}]$ GeV, the
progress of leptogenesis can be described by two-flavor approximation. The lepton asymmetries 
generated in this region can be quantified in terms of $\Delta_{B/3-L_a}$ and $\Delta_{B/3-L_\tau}$.
The muon Yukawa coupling is out of equilibrium for temperature $T > 10^9$ GeV. Below this temperature, leptogenesis has to be studied treating the three flavors separately
and the study becomes fully flavored. For our purpose, we only consider two-flavor approximation. 

 The out of equilibrium triplet decays give rise to $B/3-L_{i}$ and $\phi$ asymmetries. The $\phi$ asymmetry gets partially transferred to SM fermion asymmetries, mediated by Yukawa interactions. To analyze the evolution of $\phi$ asymmetry, no  separate Boltzmann equation is needed as its evolution is determined by the chemical equilibrium condition carried out by the reactions in thermal equilibrium at a specific temperature. The triplet abundancy relies on the reactions that either generate or wash out the $B/3-L_{i}$ asymmetry in the hot plasma. There are four classes of such interactions namely,
 \begin{enumerate}
  \item Yukawa and scalar induced decays and inverse decays 
  ($\Delta\rightarrow \bar{L} \bar{L}$ and $\Delta\rightarrow
  \phi \phi$) with total decay rate density is given by,
  $\gamma_D = \gamma_L + \gamma_\phi$.
 \item Gauge induced $2\leftrightarrow2$ scatterings, $\gamma_A$.  
  \item Lepton flavor and lepton number violating $s$- and $t$-channel $2\leftrightarrow2$ 
  scatterings mediated by triplet scalars and induced by Yukawa type interactions ($\phi \phi \leftrightarrow \bar{L}_i \bar{L}_j$ and $\phi L_j \leftrightarrow \bar{\phi} \bar{L}_i$) with reaction densities, $\gamma^{\phi \phi}_{L_i L_j}$ and $\gamma^{\phi L_j}_{\phi L_i}$.
  \item Lepton flavor violating $s$- and $t$-channel $2\leftrightarrow2$  scatterings mediated by triplets 
  ($L_n L_m \leftrightarrow L_i L_j$ and $L_j L_m \leftrightarrow
  L_i L_n$) with reaction densities, $\gamma^{L_n L_m}_{L_i L_j}$ and $\gamma^{L_i L_m}_{L_i L_n}$.
  \end{enumerate}
Among these interactions, considering only decays and inverse decays, gauge induced reactions and the off-shell pieces of the $s$-channel process can lead to a quite accurate calculation of $B-L$ asymmetry.
The Boltzmann equations relevant to flavored leptogenesis are given by \cite{Sierra:2014tqa},
\begin{equation}
 szH(z)\frac{d\Delta_{\Delta}}{dz}=-\gamma_{D}\left[\frac{\Delta_{\Delta}}{\Sigma^{eq}_{\Delta}}-\sum_{k}\left(\sum_{i}B_{L_{i}}C^{l}_{ik}-B_{\phi}C^{\phi}_{k}\right)\frac{\Delta_{k}}{Y^{eq}_{L}}\right],
 \label{eps1}
\end{equation}
\begin{equation}
 szH(z)\frac{d\Sigma_{\Delta}}{dz}=-\left[\frac{\Sigma_{\Delta}}{\Sigma^{eq}_{\Delta}}-1\right]\gamma_{D}-2\left[\frac{\Sigma^{2}_{\Delta}}{\Sigma^{eq^{2}}_{\Delta}}-1\right]\gamma_{A},
 \label{eps2}
 \end{equation}
\begin{eqnarray}
\nonumber
 szH(z)\frac{d\Delta_{B/3-L_{i}}}{dz} &=&-\left[\frac{\Sigma_{\Delta}}{\Sigma^{eq}_{\Delta}}-1\right]\epsilon^{L_{i}}\gamma_{D}+2\sum_{j}\left[\frac{\Delta_{\Delta}}{\Sigma^{eq}_{\Delta}}-\frac{1}{2}\sum_{k}C^{l}_{ijk}\frac{\Delta_{k}}{Y^{eq}_{L}}\right]B_{L_{ij}}\gamma_{D} \\ \nonumber
 & & -2 \sum_{j,k} \left(C_k^\phi + \frac{1}{2} C_{ijk}^l\right) \frac{\Delta_k}{Y^{eq}_L}
 \left( \gamma'^{\phi \phi}_{L_i L_j}+\gamma^{\phi L_j}_{\phi L_i}\right)\\ 
 & & - 
 \sum_{j,m,n,k} C^l_{ijmnk} \frac{\Delta_k}{Y^{eq}_L} \left( \gamma'^{L_n L_m}_{L_i L_j} +
 \gamma^{L_m L_n}_{L_i L_n}\right).
 \label{eps3}
\end{eqnarray}
where $z=\frac{M_{\Delta}}{T}$,
$H(z)$ is the Hubble's rate of expansion of the Universe, $H(z)=\frac{H_{0}(M_{\Delta})}{z^{2}}$, $H_{0}(T)=\sqrt{\frac{4\pi^{3}}{45}g_{*}}\frac{T^{2}}{M_{P}}$, $M_{P}=1.22\times 10^{19}$ GeV is the Planck mass, $s=(\frac{2\pi^{2}}{45})g_{*}T^{3}$ is the total entropy density and $g_{*}=106.75$. The asymmetry, $\Delta_{x}=\frac{n_{x}-n_{\overline{x}}}{s}$ where $n_x(n_{\bar{x}})$ is the number density of the species, $x(\bar{x})$, and $\Sigma_{\Delta}=\frac{n_{\Delta}+n_{\overline{\Delta}}}{s}$ is the total triplet number density. Also, $Y_x=\frac{n_{x}}{s}$ is the comoving number density, 
\begin{equation*}
 Y^{\rm eq}_{\Delta}=\frac{45g_{\Delta}}{4\pi^{4}g_{*}}z^{2}K_{2}(z), \quad
 Y^{\rm eq}_{L}=\frac{3}{4}\frac{45\varsigma(3)}{2\pi^{4}g_{*}}g_{L}, \quad
 Y^{\rm eq}_{\phi}=\frac{45\varsigma(3)}{2\pi^{4}g_{*}}g_{\phi},
\end{equation*}
are the equilibrium values, with $g_{\Delta}=1$ for each triplet component, $g_{L}=2$ and $g_{\phi}=2$, $\varsigma(3)\simeq1.202$, $K_2(z)$ is the modified Bessel function. 
The decay rate densities of triplet and gauge scattering processes are given in the (\ref{app:rates}). The components of the asymmetry vector are given by,
\begin{equation*}
 \Delta_{k}=
 \begin{pmatrix}
  \Delta_{\Delta}\\
  \Delta_{B/3-L_{a}}\\
  \Delta_{B/3-L_{\tau}}
 \end{pmatrix}.
\end{equation*}
The matrices $C^l_{ijk}$ and $C^l_{ijmnk}$
are defied as
\begin{equation}
 C^l_{ijk}=C^{l}_{ik}+C^{l}_{jk}, \quad
C^l_{ijmnk} =  C^{l}_{ik}+C^{l}_{jk} - C^{l}_{mk} - C^{l}_{nk},
\end{equation}
where the asymmetry coupling matrices, $C^l$ and $C^\phi$ are determined by the constraints coming from the chemical equilibrium condition and global symmetries of the effective Lagrangian. The matrices relate the asymmetry in lepton and scalar doublets with the $B/3 -L_i$ and triplet asymmetries as,
\begin{equation}
 \Delta_{B/3 - L_{i}}=-\sum_{k}C^{l}_{ik}\Delta_{k}, \quad 
 \Delta_{\phi}=-\sum_{k}C^{\phi}_{k}\Delta_{k}.
 \label{eq:l-phi-k}
\end{equation}
The final baryon asymmetry is estimated according to  $\eta_B = 7.04 ~ \Delta_B$, where
\begin{equation}
 \Delta_{B}=3\times \frac{12}{37}\sum_{i}\Delta_{B/3-L_{i}}.
 \label{eq:etaB-flav}
\end{equation}
 Before solving the flavored Boltzmann equations, we need to write down the relevant equations for the temperature regime where the quantum coherence of lepton flavors is restored. There are two different temperature regimes corresponding to the latter: where top quark Yukawa interactions are either fast ($10^{12}$ GeV $\lesssim T \lesssim 10^{15}$ GeV) or slow ($T \gtrsim 10^{15}$ GeV). 
 They are often dubbed as one flavor or single flavor or unflavored regime. The appropriate set of equations are given by, 
 \begin{equation}
 sHz\frac{d\Sigma_\Delta}{dz}=-\left(\frac{\Sigma_\Delta}{\Sigma^{eq}_\Delta}-1\right)\gamma_D-2\left(\frac{\Sigma^2_\Delta}{\Sigma^{2eq}_\Delta}-1\right)\gamma_A,
 \end{equation}
 \begin{equation}
 sHz\frac{d\Delta_\Delta}{dz} =-\gamma_D\left(\frac{\Delta_\Delta}{\Sigma^{eq}_\Delta}+ \sum_k\left( B_L C_k^l- B_\phi C^\phi_k\right) \frac{\Delta_k}{Y^{eq}_L}\right),
\end{equation}
 \begin{eqnarray}
 \nonumber
 sHz\frac{d\Delta_{B-L}}{dz} &=&-\left(\frac{\Sigma_\Delta}{\Sigma^{eq}_\Delta}-1\right)\gamma_D\epsilon+2\gamma_D B_L\left(\frac{\Delta_\Delta}{\Sigma^{eq}_\Delta}-\sum_k C^l_k\frac{\Delta_k}{Y^{eq}_L}\right)\\ \nonumber
 &-& 2 \sum_k 
 \left( C^\phi_k + C^l_\phi \right) 
 \frac{\Delta_k}{Y^{eq}_L} \left(
 \gamma'^{\phi \phi}_{LL} + \gamma^{\phi L}_{\phi L}\right),
 \end{eqnarray}
where
\begin{equation}
 \Delta_k =
 \begin{pmatrix}
  \Delta_\Delta\\
  \Delta_{B-L}
 \end{pmatrix}.
\end{equation}
The relation between the lepton doublet asymmetry and $\Delta_k$ is similar to the relation given in Eq.(\ref{eq:l-phi-k}), dropping the lepton flavor index. The above evolution equations in the unflavored regime match with 
that in Ref. \cite{Hambye:2005tk} provided all the SM Yukawa interactions are neglected. This is justified because a fairly accurate estimation of the resulting $B-L$ asymmetry can be carried out by considering only decays, inverse decays and gauge induces interactions as the reaction rates of Yukawa induced interactions are less by orders of magnitude from the former.
For unflavored regime, there is an alternate way of calculating the lepton asymmetry ($\Delta_{B-L} \equiv \Delta_L$), by solving the corresponding set of Boltzmann equations and is given by,
\begin{equation}
\Delta_L = \epsilon~ \eta~\Sigma_\Delta\rvert_{z \ll 1},
\label{eq:effi}
\end{equation}
which can be partially converted to baryon asymmetry by sphalerons,
\begin{equation}
 \eta_B = -0.029 ~ \epsilon ~ \eta,
 \label{eq:eta_b}
\end{equation}
where $\eta$ is the efficiency factor, can be obtained by solving the complete set of unflavored Boltzmann equations.
\section{Quantitative analysis of Baryon asymmetry}
\label{sec:quant-bau}
In this section, we calculate the final BAU from the decay of the lightest of the two triplets. In our case, it is $\Delta_2 \equiv \Delta$. First, we consider the unflavored or single flavored regime and find that the scale of baryogenesis through leptogenesis can be as low as $1 \times 10^{10}$ GeV. As discussed in the previous section, the predicted scale is suitable for study of flavored (two-flavored) leptogenesis. We take up this study consequently. In both the scenarios,
first we estimate the amount of CP asymmetry in terms of the model parameters and then using the latter, we solve the appropriate set of Boltzmann equations.
\subsection{Baryogenesis through unflavored leptogenesis}
\label{sec:unflav}
In this case, the lepton flavors act as single entity and are indistinguishable. The CP asymmetry parameter
now can be expressed in line with Eq.(\ref{eps}) as,
\begin{equation}
 \epsilon=\frac{2M_{\Delta}}{4\pi v^2}(B_L B_{\phi})^{1/2}\frac{{\rm Im} [{\rm Tr}(m^{(2)}_{\nu}m^{(1)\dag}_{\nu})]}{[Tr(m^{(2)\dag}_{\nu}m^{(2)}_{\nu})]^{\frac{1}{2}}}. 
 \label{eq:branco}
\end{equation}
To get the upper bound on the CP asymmetry the above equation
can be written in terms of inequality as \cite{Hambye:2005tk},
\begin{eqnarray}
\nonumber
 |\epsilon| &\lesssim & \frac{2 M_\Delta (B_L B_\phi)^{1/2}}
 {4\pi \upsilon^2} \left[ {\rm Tr} (m_\nu^\dagger m_\nu) \right]^{1/2}\\
 &=& \frac{2 M_\Delta (B_L B_\phi)^{1/2}}{4\pi \upsilon^2}  \left( \sum_{k =1}^3 m_k^2\right)^{1/2}.
\end{eqnarray}
For hierarchical light neutrinos, the upper bound on the CP asymmetry parameter is given by,
 \begin{equation}
  |\epsilon|\lesssim10^{-6}\times(B_{L}B_{\phi})^{\frac{1}{2}}\frac{M_{\Delta}}{10^{10}GeV}\frac{m_{atm}}{0.05eV}.
 \end{equation}
It is clear to see that the CP asymmetry is maximum for 
$B_{L}= B_\phi = 0.5$. But as we proceed further, it becomes clear that by solving the full set of Boltzmann equations, for equal branching ratios the efficiency is suppressed. Rather with strongly hierarchical branching ratios ($B_L\gg B_\phi$ or $B_\phi \gg B_L$) it is possible
to get maximal efficiency. But the latter choice would result in suppressed CP asymmetry. Nevertheless, we find with strongly hierarchical branching ratios it is possible to get
sizable CP asymmetry from the contribution of the Majorana phases and phase of the triplet vev, $\alpha$. Also requiring successful leptogenesis from decay of the triplet scalar one can put some constraints on the phase, $\alpha$.

For a realistic calculation in our case, by using Eq.s (\ref{nN}) in eq.(\ref{eq:branco}) and  we obtain,
\begin{equation}
 \epsilon=\frac{2M_{\Delta}}{4\pi v^2}(B_LB_{\phi})^{1/2}F(\alpha,\phi,\omega_i) \sqrt{A^{2}+2B^{2}+2C^{2}+D^{2}+2M^{2}+N^{2}},
 \label{eq:unflav-cp}
\end{equation}
where
\begin{equation}
 F(\alpha,\phi,\omega_i) = \frac{|\omega_{1}|\sin(\phi-\alpha)}{\sqrt{f(|\omega_{1}|, \omega_{2}, \alpha, \phi)}};~ f(|\omega_{1}|, \omega_{2}, \alpha, \phi)=|\omega_{1}|^{2}+\omega^{2}_{2}+2|\omega_{1}|\omega_{2} \cos(\phi-\alpha).
\end{equation}
In order to get the maximal CP asymmetry, we have made an assumption in the preceding calculation. We have assumed that the phase differences of the $ij$-th elements of $Y_\Delta^{(1)}$ and $Y_\Delta^{(2)}$
conspire to be equal; $\phi_{a'}-\phi_{a} = \phi_{b'}-\phi_{b}= \phi_{c'}-\phi_{c}=\phi_{d'}-\phi_{d}=\phi_{m'}-\phi_{m}= \phi_{n'}-\phi_{n} = \phi$. 
From Eq.(\ref{eq:beta}) the value of $\alpha$ is constrained to be $\alpha \neq n\pi, ~n=0,1,2,3,\cdots.$
To ensure non-zero CP violation, $\phi \neq \alpha$, which  is compatible with our assumption that the phases in the $ij$-th elements of $Y_\Delta^{(1)}$ and $Y_\Delta^{(2)}$ are necessarily different. For particular benchmark points the values of triplet 
Yukawa couplings are presented in the  \ref{sec:elements}.
The parameters, $A,B,C,D,M$ and $N$ are determined from neutrino oscillation data, as described in section (\ref{sec:trip-model}). 
In calculating the values of $\epsilon$, the following values of the elements of the neutrino mass matrix are taken.
\begin{eqnarray}
\nonumber
 A &=& 0.028306~{\rm eV}, ~ B=0.012501~{\rm eV}, ~ C= 0.015691~{\rm eV}, \\ 
 D &=& 0.018767~{\rm eV}, ~ M = 0.017914~{\rm eV}, ~ N= 0.016304~{\rm eV}.
\end{eqnarray}
The preceding values of the parameters are chosen from Fig.(\ref{fig:mdnplots}) for a given value of $m_1$ in such a way that 
the values of the light neutrino masses for normal hierarchy satisfy the recent Planck data \cite{Planck:2018vyg}. They are given by,
\begin{equation}
 m_{1}=0.030000 ~{\rm eV}, \quad
 m_{2}=0.031131 ~{\rm eV}, \quad  
 m_{3}=0.058134 ~{\rm eV},
 \end{equation}
and the sum of neutrino masses,
\begin{equation}
 \Sigma=m_{1}+m_{2}+m_{3}=0.119266 ~{\rm eV},
\end{equation}
is bounded by Planck data.
Here we have not taken the renormalization group evolution of neutrino mass matrix into account. As shown in Ref. \cite{Hambye:2005tk}, the result is only
$m_\nu (\rm high ~scale) \sim (1.2/1.3)~ m_\nu$(low scale) \cite{Barbieri:1999ma, Giudice:2003jh}.
\begin{figure*}[t]
\includegraphics[width=0.52\textwidth,height=0.38\textwidth]{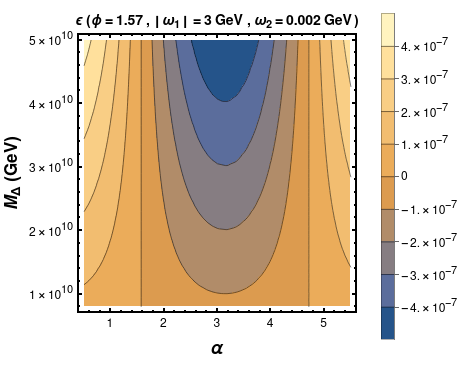}\hspace{0.158cm}
\includegraphics[width=0.52\textwidth,height=0.38\textwidth]{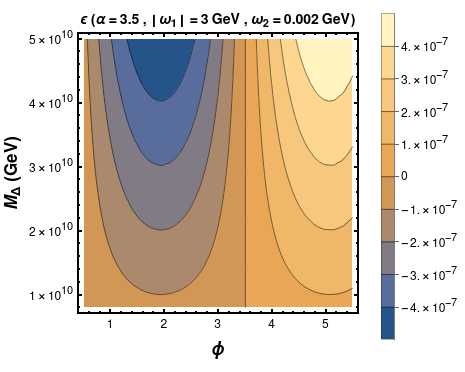} 
\vspace{-0.2cm}
\caption{Contour plot of CP asymmetry parameter $\epsilon$ as a function of the phases $\alpha$ and $\phi$ and triplet mass $M_{\Delta}$. In the left panel, the contours of $\epsilon$ are plotted for phase $\phi=1.57$ by varying triplet mass and $\alpha$ in the range $\left[8\times10^{9},5\times 10^{10}\right]$ GeV and $[0, 2\pi]$ respectively. In the right panel the contours of $\epsilon$ are plotted for phase $\alpha=3.5$ by varying triplet mass and $\phi$ in the range $\left[8\times10^{9},5\times 10^{10}\right]$ GeV and $[0, 2\pi]$ respectively.
The triplet vevs $|\omega_{1}|$ and $\omega_{2}$ are kept fixed at $3$ GeV and $0.002$ GeV, respectively. }
\label{fig:alpha-epsilon}
\end{figure*}

\begin{figure*}[h!]
\includegraphics[width=0.49\textwidth,height=0.38\textwidth]{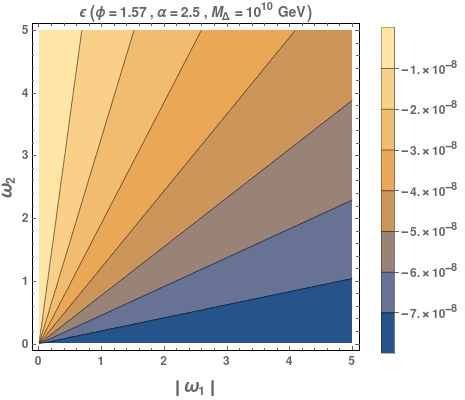}\hspace{0.158cm}
\includegraphics[width=0.49\textwidth,height=0.38\textwidth]{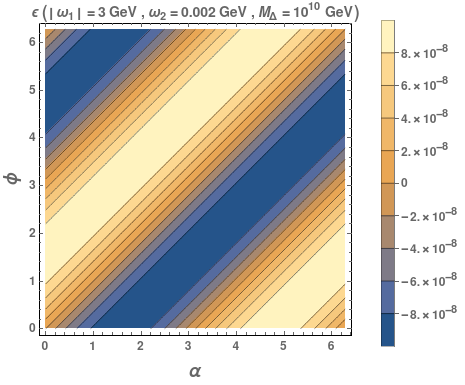} 
\vspace{-0.2cm}
\caption{Contour plots of CP asymmetry parameter, $\epsilon$ as a function of the triplet vevs, $|\omega_{1}|$ and $\omega_{2}$, and phases $\alpha$ and $\phi$ for triplet mass $M_{\Delta}=1\times10^{10}$ GeV. The values of phases are kept fixed at $\phi=1.57$ and $\alpha=2.5$. In the left panel the contours of $\epsilon$ are plotted for $\alpha =2.5$ and $\phi =1.57$. In the right panel, contours of $\epsilon$ are plotted as a function of the phases $\alpha$ and $\phi$ and the triplet vevs are kept fixed at $|\omega_{1}|=3$ GeV and $\omega_{2}=0.002$ GeV, respectively.}
\label{fig:omega-alpha-phi}
\end{figure*}

To get baryon asymmetry $\sim \mathcal{O}(10^{-10})$, the $\epsilon$
parameter should be of the order $\mathcal{O}(10^{-6} - 10^{-8})$ requiring efficiency,
$\eta$ of the order $\mathcal{O}(10^{-3} - 10^{-2})$. 
The CP asymmetry parameter in Eq.(\ref{eq:unflav-cp}) depends on the parameters: $M_\Delta$, the two phases
$\alpha$ and $\phi$, apart from the branching ratios $B_L$ and $B_\phi$. 
We plot correlation plots among $\epsilon$, $M_\Delta$, $\phi$, $\alpha$ and vevs of the triplet scalars, $|\omega_1|$ and $\omega_2$. They are shown in the figures (\ref{fig:alpha-epsilon}) and (\ref{fig:omega-alpha-phi}). In Fig. (\ref{fig:alpha-epsilon}) the values of the triplet vevs are taken to be
$|\omega_1| = 3$ GeV and $\omega_2 = 0.002$ GeV. 
We have taken $B_L =0.995$. In the left panel of 
Fig.(\ref{fig:alpha-epsilon}), it shows that adequate CP violation, $\epsilon$ ($\sim 10^{-7}$) can be produced for triplet mass as low as 
$1\times 10^{10}$ GeV for $\phi =1.57$ and for a range of values of $\alpha$. In the right panel of the same figure, it is shown that adequate CP violation can be generated for triplet mass as low as 
$1\times 10^{10}$ GeV for $\alpha= 3.5$ and for a
range of values of $\phi$. This result saturates the bound of order of magnitude of $M_\Delta$ as reported by the authors of Ref. \cite{Hambye:2005tk}. In Fig.(\ref{fig:omega-alpha-phi}), in the 
left panel, regions of parameter space of $|\omega_1|$ and $\omega_2$ are shown for generation of adequate CP violation for 
$M_\Delta = 10^{10}$ GeV and $\phi = 1.57$ and $\alpha = 2.5$.
Similarly, in the right panel, the parameter space of $\alpha$ and
$\phi$ is shown for $M_\Delta = 10^{10}$ GeV, $|\omega_1| = 3$ GeV and $\omega_2 = 0.002$ GeV that can give rise to desired magnitude
of CP violation for generating BAU in the observational range.

\begin{figure*}[h!]
\includegraphics[width=0.49\textwidth,height=0.38\textwidth]{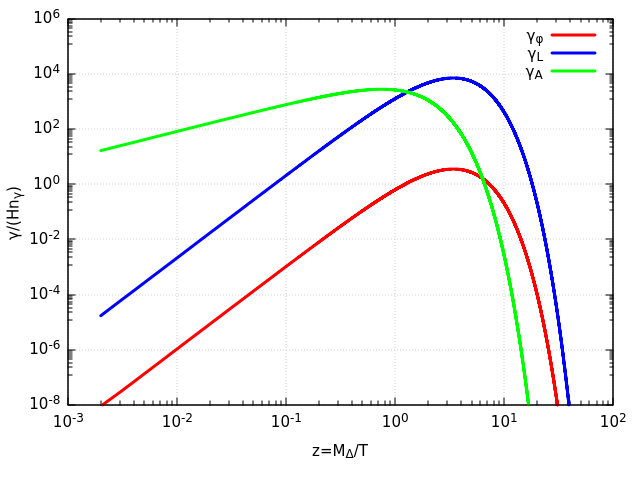}\hspace{0.158cm}
\includegraphics[width=0.49\textwidth,height=0.38\textwidth]{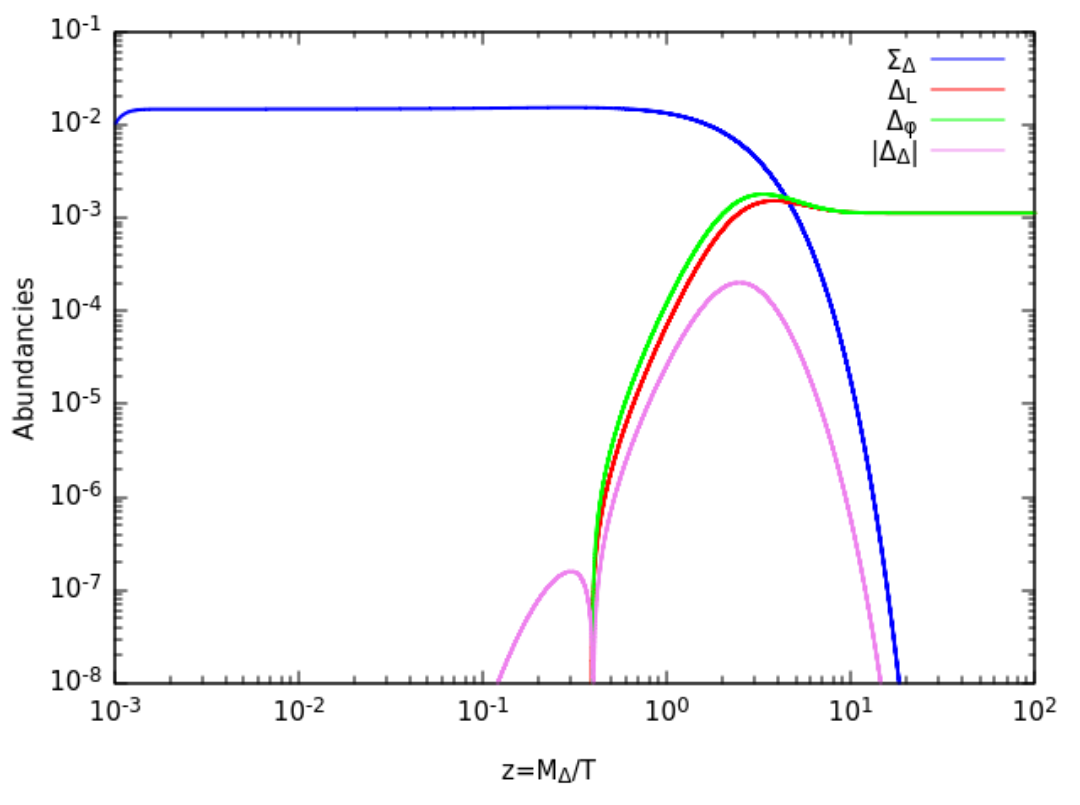} 
\vspace{-0.2cm}
\caption{ In the left panel the figure shows the evolution of interaction rates $\gamma_{\phi}, \gamma_{L}$ and $\gamma_A$ in units of $H n_\gamma$ with $z$. The quantities $\gamma_{\phi}$ and $\gamma_{L}$ represent space-time densities of decay rates  of the decay of triplets to the SM Higgs and leptons respectively. The space-time density
of gauge scattering of the triplets are represented as $\gamma_{\phi}$. In the right panel the figure depicts the evolution of abundancies with $z$ 
for $B_{L}=0.9995,B_{\phi}=0.0005$.}
\label{fig:blbh11}
\end{figure*}

Next, we solve Boltzmann equations as given in 
Ref. \cite{Hambye:2005tk} to estimate the efficiency of leptogenesis. The equations are also given in (\ref{app:rates}).  
 The relevant abundancies are plotted in Fig.(\ref{fig:blbh11}) in units of $|\epsilon|$ against $z=\frac{M_{\Delta}}{T}$ for hierarchical $B_L$ and $B_\phi$. In case of triplet scalar leptogenesis, the efficiency factor can be defined as the ratio of lepton asymmetry $\Delta_{L}$, in units of CP asymmetry parameter $\epsilon$ and the total triplet density $\Sigma_{\Delta}$ for $z\ll1$,
 \begin{equation}
  \eta=\frac{\Delta_{L}/\epsilon}{\Sigma_{\Delta}(z\ll1)}.
  \label{etab}
 \end{equation}

In Fig.(\ref{fig:blbh11}), $\gamma_{L}=B_L\gamma_D$, $\gamma_{\phi}=B_{\phi}\gamma_D$ and $\gamma_{A}$ are plotted against $M_{\Delta}/T$. For this, we  take the value of $B_L= 0.9995$. The numerical results for other choices of $B_L (= 0.995, 0.5)$ are also discussed. 
The decay rates $\gamma_{L}$, $\gamma_{\phi}$ and $\gamma_{A}$ are shown for different values of branching ratios, $B_{L}$, $B_{\phi}$ in Fig.(\ref{fig:blbp}). The corresponding abundancy plots are also shown therein. It is to be noted that the efficiency factor is proportional to the asymmetry $\Delta_{L}$, as expressed in Eq.(\ref{etab}), and it is observed from the figures (\ref{fig:blbh11}) and (\ref{fig:blbp}) that, when $B_{L}\gg B_{\phi}$ or $B_{\phi}\gg B_{L}$, the asymmetry $\Delta_{L}$ is larger as compared to the
case of $B_L = B_\phi$. Hence, the efficiency is found to be maximal in the former case. This feature emerges because, the washout is negligible in the decay channel with small branching ratios. On the other hand, the reduced asymmetry $\Delta_{L}$ makes efficiency minimal when $B_{L}=B_{\phi}=0.5$.  

\begin{table}
 \centering 
 \begin{tabular}{ |p{2.0cm}||p{2.0cm}||p{2.5cm}||p{2.5cm}||p{3.0cm}|  }
 \hline
 $M_{\Delta}({\rm GeV})$&$B_{L}$&$|\epsilon| $&$\eta$&$\eta_{B}$\\
 \hline
 $1\times10^{10}$&$0.995$&$8.0\times10^{-8}$&$0.021445$&$4.975\times10^{-11}$\\
 $1\times10^{10}$&$0.9995$&$3.0\times10^{-8}$&$0.066$& ${\bf 5.742\times10^{-11}}$\\
 $5\times10^{10}$&$0.995$&$4.0\times10^{-7}$&$0.0240$&$2.784\times10^{-10}$\\
 $5\times10^{10}$&$0.9995$&$1.5\times10^{-7}$&$ 0.096878$&$ 4.214\times10^{-10}$\\
 $1\times10^{11}$&$0.995$&$8.0\times10^{-7}$&$0.01675$&$3.886\times10^{-10}$\\
 $1\times10^{11}$&$0.9995$&$3.0\times10^{-7}$&$0.1$&$8.7\times10^{-10}$\\
 \hline
\end{tabular}
\caption{Baryon asymmetries obtained for different values $M_\Delta, B_L$ for unflavored leptogenesis.
The values of CP asymmetry, $\epsilon$ are calculated for each case. The efficiency, $\eta$ is also calculated using eq.(\ref{etab}) for each case using the unflavored Boltzmann equations.}
\label{tab:neop}
\end{table}
For hierarchical branching ratios, the amount of CP asymmetry is suppressed but the efficiency is enhanced.  Fig.(\ref{fig:blbh11}) shows one of the case of hierarchical branching ratios with $M_\Delta = 1 \times 10^{10}$ GeV, whereas the figures (\ref{fig:blbp})  show that the efficiency decreases significantly for equal branching ratios. We present some of the instructive numerical estimates of CP violation parameter, efficiency and amount of BAU in table (\ref{tab:neop}).
We can summarise the outcomes of this section as follows. As it can be seen from table(\ref{tab:neop}),
for a given value of $M_\Delta$ the CP asymmetry parameter, $\epsilon$ decreases with the increase in the hierarchy among branching ratios, $B_L$ and $B_\phi$. But the efficiency increases with increase in hierarchy among branching ratios, so also the baryon asymmetry.
For example, for $M_\Delta = 5\times 10^{10}$ GeV, 
$\epsilon$ decreases from $4.0 \times 10^{-7}$ to 
$1.5\times10^{-7}$ by increasing $B_L$ from $0.995$ to $0.9995$. Whereas efficiency, $\eta$ increases
from $0.024$ to $0.096878$. The baryon asymmetry increases
from $2.784\times 10^{-10}$ to $4.214\times 10^{-10}$.
The results are similar for $B_\phi \gg B_L$. 
Nevertheless, a maximal value of $\epsilon$ ($\sim 10^{-7}$) can be achieved for triplet mass as low as 
$1 \times 10^{10}$ GeV due to the contributions 
of phases, $\phi$ and $\alpha$ even for hierarchical
branching ratios. Such a value of triplet mass
offers the motivation to study baryogenesis through leptogenesis considering the flavor effects which we take up in the next section. 

\subsection{Baryogenesis through flavored Leptogenesis}
\label{sec: Flav1}
Drawing insights from the previous section of unflavored leptogenesis (table (\ref{tab:neop})) we take $M_\Delta = 1 \times 10^{10}$ GeV. This mass range fits suitably to study leptogenesis in two flavored regime, as discussed in section (\ref{sec:BE-discuss}). Here, first we calculate the relevant CP asymmetry parameters using 
Eq.s (\ref{eq:CP-tot}), (\ref{eq:LF}) and (\ref{eq:F}) and the parameters are suitably taken from neutrino data as explained in section (\ref{sec:trip-model}). Here the total CP
asymmetry parameter is calculated from two contributions,
$\epsilon^\tau$ and $\epsilon^a$, where $\epsilon^a = \epsilon^e+\epsilon^\mu$, as in this regime there are
only two distinct flavors of leptons; $\tau$ and $a$. The latter is a linear combination of $e$ and $\mu$ flavors.

Before we calculate the total CP asymmetry from the decay of the triplet scalars, some comments are in order. In case of unflavored or single flavored leptogenesis, we see that for  hierarchical branching ratios $(B_L \ll B_\phi ~\rm{or}~ B_\phi \ll B_L)$, though the CP asymmetry is suppressed, the efficiency is maximal. As discussed in the section (\ref{sec:CPasym-unflav}), the parameter $\epsilon$ gets contribution from two sources as is given in Eq.(\ref{eq:CP-tot}). The second contribution on the right side of the Eq.(\ref{eq:CP-tot}) dominates over the first one if triplets couple more to leptons than Higgs doublet. In other words, this corresponds to the case
$B_L \gg B_\phi$ and the scenario is called PFL. The  tree level branching ratios of the triplet scalar to leptons are given in Eq. (\ref{eq:br-ratio}).
The tree level branching ratios of the triplet scalar to leptons in the two flavored regime is given as,
\begin{equation*}
 B_{L}=\sum_{i,j=a,\tau}B_{L_{ij}},
 \label{eq:Bl}
\end{equation*}
where
\begin{equation}
B_{L_{aa}} = \frac{M_{\Delta}}{8\pi\Gamma_{\Delta}}\sum_{i,j=1,2}|Y^{(2)}_{\Delta_{ij}}|^{2}, ~
 B_{L_{\tau a}} = \frac{M_{\Delta}}{8\pi\Gamma_{\Delta}}\sum_{j=1,2}|Y^{(2)}_{\Delta_{3j}}|^{2},~ B_{L_{\tau\tau}}=\frac{M_{\Delta}}{8\pi\Gamma_{\Delta}}|Y^{(2)}_{\Delta_{33}}|^{2}. 
 \label{eq:Bl-ij}
\end{equation}
In our study, the determination of Yukawa parameters ($Y_{\Delta_{ij}}$'s) are relevant for determination of leptonic branching ratios. They are also relevant for calculating  the CP asymmetry parameter as given in Eq.(\ref{eq:CP-tot}), and the following Eq.s (\ref{eq:LF}) and (\ref{eq:F}). So, both of the quantities are determined from $Y_{\Delta_{ij}}$'s through neutrino mass matrix elements $A, B, C, D, M, N$ and the phases, as can be seen from Eq.s (\ref{eq:conjecture}) and (\ref{eq:A-to-N}).  
We read the values of $A ~{\rm to}~ N$ from fig.(\ref{fig:mdnplots}).  

We calculate CP asymmetry parameter by using Eq.s (\ref{eq:CP-tot}), (\ref{eq:LF}) and (\ref{eq:F}) and we further substitute, $\phi_{a'}-\phi_{a}=\phi_{1}$, $\phi_{b'}-\phi_{b}=\phi_{2}$, $\phi_{c'}-\phi_{c}=\phi_{3}$, $\phi_{d'}-\phi_{d}=\phi_{4}$, $\phi_{m'}-\phi_{m}=\phi_{5}$, $\phi_{n'}-\phi_{n}=\phi_{6}$, then we define a function $f_{i}(|\omega_{1}|,\omega_{2},\alpha,\phi_{i})$ as given by,
\begin{equation*}
f_{i}=|\omega_{1}|^{2}+\omega^{2}_{2}+2|\omega_{1}|\omega_{2}\cos(\phi_{i}-\alpha).
\label{eq:fi}
\end{equation*}
The CP asymmetry parameters take the form,
\begin{eqnarray}
\nonumber
   \epsilon^a &=& \frac{2 M_{\Delta}(B_{L}B_{\phi})^{\frac{1}{2}}|\omega_{1}|}{4\pi v^{2} F} \times \\ \nonumber
   & & \left( 
    \frac{A^{2} \sin(\phi_{1}-\alpha)}{f_1} +\frac{2B^{2} \sin(\phi_{2}-\alpha)}{f_2}
+\frac{C^{2} \sin(\phi_{3}-\alpha)}{f_3} 
+ \frac{D^{2} \sin(\phi_{4}-\alpha)}{f_4}+ \frac{M^{2} \sin(\phi_{5}-\alpha)}{f_{5}}
\right),\\
\epsilon^\tau &=& \frac{2 M_\Delta (B_{L}B_{\phi})^{\frac{1}{2}}|\omega_{1}|}{4\pi v^{2} F} \times 
  \left(
  \frac{C^{2} \sin(\phi_{3}-\alpha)}{f_{3}}+\frac{M^{2} \sin(\phi_{5}-\alpha)}{f_5}
  + \frac{N^{2} \sin(\phi_{6}-\alpha)}{f_6}\right).
\label{eq:epsilon-a-tau}
\end{eqnarray}
where $F=\sqrt{\frac{A^{2}}{f_{1}}+\frac{2B^{2}}{f_{2}}+\frac{2C^{2}}{f_{3}}+\frac{D^{2}}{f_{4}}+\frac{2M^{2}}{f_{5}}+\frac{N^{2}}{f_{6}}}$. The above equations can be obtained 
in terms of neutrino mass matrix elements as given in Ref. \cite{Branco:2011zb}. Like the case of unflavored leptogenesis here also we see that the CP asymmetry parameters, $\epsilon^a$ and $\epsilon^\tau$ vanish for $\phi_i = \alpha, i =1, 2, \cdots  6$. The non-vanishing CP asymmetry  ensures the phases, $\phi_i \neq m \pi, m = 0, 1, 2, \cdots$.
We vary the phase values randomly in $(0- 2\pi)$ range and we choose sets 
of phase values that give maximum CP asymmetry.
We choose two sets of the such values and they are given by,
 \begin{itemize}
  \item Set-I\\
 In this set, we choose $\alpha=2.90$, $\phi_{1}=\phi_{2}=\phi_{3}=\phi_{4}=1.5$, $\phi_{5}=\phi_{6}=1.30$.
 \item Set-II\\
 In this set, we choose $\alpha=2.50$, $\phi_{1}=\phi_{2}=\phi_{3}=\phi_{4}=1.5$, $\phi_{5}=1.70$, $\phi_{6}=5.09$.
\end{itemize}
In checking the individual values
of $B_{L_{ij}}$'s and $\epsilon^{L_i}$'s for all the
random sets of phase values we find the contribution of purely flavored part, $\epsilon^{L_i (\cancel{F})}$ to the flavored CP asymmetry parameter, $\epsilon^{L_i}$ comes out negligibly small. Although with that particular 
choice of phases the value of branching ratio, $B_L$ 
as estimated in Eq.(\ref{eq:Bl-ij}) comes maximal. For illustration, we take the values given in Set II
and using Eq.(\ref{eq:Bl-ij}), taking  
$M_{\Delta}=1\times10^{10}$GeV, we find $B_{L_{aa}}=0.51118$, $B_{L_{a\tau}}=B_{L_{\tau a}}=0.197728$, $B_{L_{\tau\tau}}= 0.092863$ and $B_{L}=0.9995$. The purely flavor violating CP asymmetry parameters come out to be $\epsilon^{\tau(\cancel{F})}\sim4.336\times 10^{-42}$ and $\epsilon^{a(\cancel{F})}\sim-4.337\times 10^{-42}$, where $\epsilon^{a(\cancel{F})}=\epsilon^{e(\cancel{F})}+\epsilon^{\mu(\cancel{F})}$, $\epsilon^{e(\cancel{F})}\sim0.386\times 10^{-42}$, $\epsilon^{\mu(\cancel{F})}\sim-4.723\times 10^{-42}$.  
 The contribution from $\epsilon^{L_i (\cancel{F})}$ for each flavor comes out vanishingly small as compared to the contribution, $\epsilon^{L_i (\cancel{L}, \cancel{F})}$. 
 
 So in our case PFL scenario is not achievable. The CP asymmetry parameter gets contribution from the lepton number and flavor violating part, $\epsilon^{L_i(\cancel{L}, \cancel{F})}$, which we calculate here. In our case, the hierarchy $B_\phi \gg B_L$ remains viable for 
successful baryogenesis through leptogenesis.

\begin{figure*}[h!]
\centering
 \includegraphics[width=0.46\textwidth,height=0.38\textwidth]{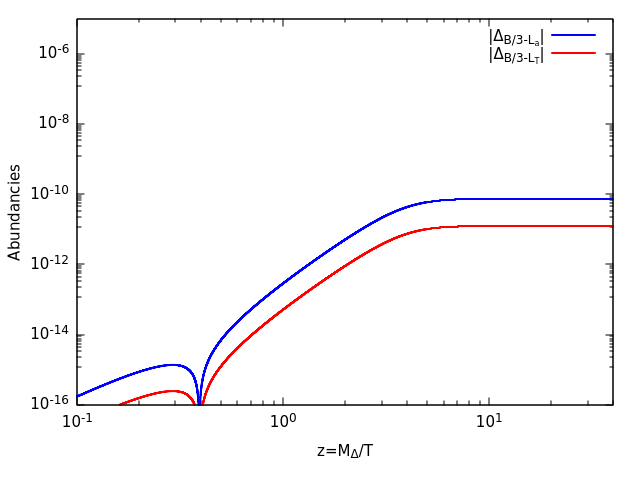}\hspace{0.5cm}
  \includegraphics[width=0.46\textwidth,height=0.38\textwidth]{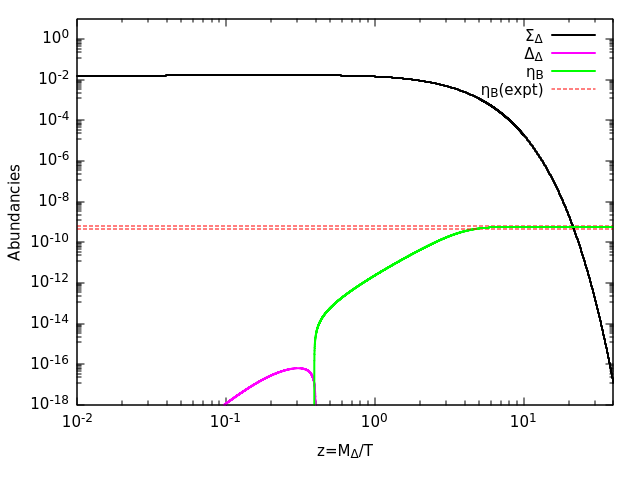}
 \caption{ In the left panel, the figure shows the evolution of the 
 lepton asymmetries for two flavors, $e$ and $a$ in two flavor approximation 
 for $M_{\Delta}=1\times10^{10}$ GeV, and $B_{\phi}=0.9995$. In the right panel,
it shows the evolution of abundancy of triplet number density and baryon asymmetry,
$\eta_B$. $\eta_B$ is calculated using eq.s (\ref{eq:l-phi-k}) and (\ref{eq:etaB-flav}) and compared with the observational value. }  \label{fig:set}
\end{figure*}

 In order to incorporate the preceding results in solving the set of flavored Boltzmann equation and finally calculating the baryon asymmetry we take two different sets of values of phases, $\phi_i$'s and $\alpha$, Set I and II. 
We take different values of $B_{\phi}$ and they are are summarised in table (\ref{tab:2flavtable1}).  Though we solve Boltzmann equations for all the cases but we show the abundances plotted in Fig.(\ref{fig:set}) only for one case that corresponds to the set of values in Set II. We take different
values of $B_\phi$ and for each case we take  two sets values and illustrate the results  for $M_{\Delta}=1\times10^{10}$ GeV. In both cases the CP violation parameters, $\epsilon^a$ and $\epsilon^\tau$ are calculated by using expressions given in the set of equations in Eq.(\ref{eq:epsilon-a-tau}).
\begin{table}
 \centering 
 \begin{tabular}{ |p{1.5cm}||p{1.5cm}||p{2.5cm}||p{2.5cm}||p{2.5cm}| }
 \hline 
 $M_{\Delta}({\rm GeV})$&$B_{\phi}$&$\epsilon^{a}$&$\epsilon^{\tau}$&$\eta_{B}$\\
 \hline
 $1\times 10^{10}$&$0.9995$&$-2.20\times 10 ^{-8}$&$-9.10\times 10^{-9}$&${\bf 8.66\times 10^{-10}}$\\
 
  $1\times 10^{10}$&$0.9995$&$-1.83\times 10^{-8}$&$-3.29\times 10^{-9}$&$\bf{5.93\times 10^{-10}}$\\
  
 \hline
 $1\times 10^{10}$&$0.995$&$-7.028\times 10^{-8}$&$-2.88\times 10^{-8}$&$1.39\times10^{-9}$\\
 
 $1\times 10^{10}$&$0.995$&$-4.06\times 10^{-8}$&$-3.77\times 10^{-9}$&$5.89\times10^{-10}$\\
 \hline
 $1\times 10^{10}$&$0.95$&$-2.06\times 10^{-7}$&$-7.46\times 10^{-8}$&$1.58\times 10^{-9}$\\
 
  $1\times 10^{10}$&$0.95$&$-8.61\times 10^{-8}$&$-1.33\times 10^{-8}$&$5.17\times 10^{-10}$\\
 \hline
\end{tabular}
\caption{Baryon asymmetries from two-flavored leptogenesis for different values of $B_{\phi}$. In each row we report two sets of values of flavored CP asymmetry parameters and the corresponding values of baryon asymmetry, $\eta_B$. The two sets of values (upper(lower) row) in the first row correspond to Set I(II) of the phases given in the main text and $B_\phi = 0.9995$. For the other reported values (for 
$B_\phi = 0.995$ and $0.95$) the phase values are chosen in a manner such that 
the value of the CP asymmetry parameter is maximal.}
\label{tab:2flavtable1}
\end{table}

For solving the set of Boltzmann equations and for the temperature under consideration the $C^{l}$ and $C^{\phi}$ matrices are given as,
\begin{equation*}
 C^{l}=
 \begin{pmatrix}
  -6/359 & 307/718 & -18/359\\
  39/359 & -21/718 & 117/359
 \end{pmatrix},
\end{equation*}
\begin{equation*}
 C^{\phi}=
 \begin{pmatrix}
  258/359 & 41/359 & 56/359
 \end{pmatrix}.
\end{equation*}
The Boltzmann equations in Eq.s (\ref{eps1} - \ref{eps3}) are solved numerically for $M_\Delta = 1 \times 10^{10}$ GeV to obtain  the lepton flavored asymmetries $\Delta_{B/3-L_i}$. Finally the baryon asymmetry is calculated by using Eq.s (\ref{eq:l-phi-k}) and (\ref{eq:etaB-flav}). The results are summarised in the
table (\ref{tab:2flavtable1}). For $B_\phi = 0.9995$ and for set of phase values in Set II, the predicted baryon asymmetry falls within observational range. Also comparing the results presented in table (\ref{tab:2flavtable1})
with that in table (\ref{tab:neop}), corresponding to $M_\Delta =1\times 10^{10}$ GeV and $B_\phi = 0.9995$, the 
baryon asymmetry is enhanced considering flavor effects
as compared to unflavored case.
\section{Low scale leptogenesis}
\label{sec:resonant}
In this section we study baryogenesis via leptogenesis
by taking the mass scale of the triplets, $M_\Delta$ close
to TeV scale. In section (\ref{sec:unflav}), we find for
hierarchical  masses of triplet scalars successful leptogenesis requires $M_\Delta \geq 1 \times 10^{10}$ GeV.
Below this bound, the CP asymmetry parameter will be highly suppressed. However, if the mass scales of the triplet scalars are quasidegenerate the leptonic asymmetry can be resonantly enhanced if the relation between
their respective masses and decay width satisfy $|M_{\Delta_j} - M_{\Delta_i}|\sim \frac{1}{2}
\Gamma_{\Delta_j}$. The CP violation parameter can be given as, after summing over the
lepton flavors \cite{Branco:2011zb},
\begin{equation}
\epsilon_{\rm resonant}\simeq\frac{\sqrt{B_{L}B_{\phi}}{\rm Im}[{\rm Tr}(m^{(2)}_{\nu}m^{(1)\dagger}_{\nu})]}{[{\rm }{\rm Tr}(m^{(2)\dagger}_{\nu}m^{(2)}_{\nu})]^{\frac{1}{2}}[{\rm Tr}(m^{(1)\dagger}_{\nu}m^{(1)}_{\nu})]^{\frac{1}{2}}}.
\label{eq:epsilon-resonant}
\end{equation}
This CP violation parameter is neither suppressed by light neutrino mass nor the scalar triplet mass. It is
 bounded by unitarity constraint, $|\epsilon| < 2 ~ {\rm min}(B_L, B_\phi)$. This feature
 opens up the possibility of triplet leptogenesis at the TeV scale. 
 
 The final BAU also depends on the triplet annihilation rate, $\gamma_A$
 through the efficiency of the leptogenesis mechanism. 
 In the energy regime ($\sim{\rm TeV}$), the triplets become non-relativistic and new features emerge in their
 annihilation rate $\gamma_A$. The non-relativistic scatterings among gauge-charged particles are
 affected by non-perturbative electroweak Sommerfeld corrections. The scatterings are distorted by the Coulomb force, as when the
 charged particles become non-relativistic, their kinetic energy is lower than the
 electrostatic potential energy. The states that are involved in the scattering are no longer plane waves. This introduces new  corrections in $\gamma_A$ and decreases the efficiency by $30\%$ \cite{Strumia:2008cf}.   Also, the 
 efficiency decreases at $M_{\Delta} \leq 3$ TeV because when the triplet decays produce lepton
 asymmetry, sphalerons no longer convert it into baryon asymmetry. And for temperature
 below the mass of the Higgs boson, the electroweak symmetry starts to break suppressing the
sphaleron transition rates. Thus a stringent bound on the
triplet mass scale arises. The baryogenesis via leptogenesis is only possible at $M_{\Delta}\geq 1.6$ TeV \cite{Strumia:2008cf}.

In the following, we briefly summarise the Sommerfeld corrections in $\gamma_A$ that finally goes into solving the Boltzmann equations to calculate the efficiency and hence final BAU.
 Referring to Eq.(\ref{eq:gamma-gen}) of the annihilation rate, the reduced cross section of
 $2 \rightarrow 2$ scatterings can be represented as
 \begin{equation}
  \hat{\sigma}_A (s) = c_s\beta + c_p\beta^3 + \mathcal{O}(\beta^5), ~\beta = \sqrt{1-4 M_\Delta^2/s},
  \label{eq:sigma-S}
 \end{equation}
 where $\beta$ is the triplet velocity in $\Delta\Delta^*$ center of mass frame.  The annihilation rate
 is given by
 \begin{equation}
  \gamma_A =  \frac{M_\Delta T^3 e^{-2M_\Delta/T}}{32\pi^3} \left[c_s +\frac{3 T}{2 M_\Delta}(c_p +\frac{c_s}{2})+
  \mathcal{O} (T/M_\Delta)^2 \right], ~ c_s = \frac{9g_2^4+12 g_2^2 g_{\mathcal Y}^2+3g_{\mathcal Y}^2}{2 \pi}
 \end{equation}
For a scalar triplet the gauge scattering density, $\gamma_A$ can be approximated by just is $s$-wave coefficient, $c_s$ (see Eq. (\ref{eq:gammaA})).
The Sommerfeld corrections modify the coefficient, $c_s$. The important sources of correction are summarised in the (\ref{app:sommer}) and one can refer to \cite{Strumia:2008cf} for detailed analysis. The modified coefficient, $c_s$ is given by,
\begin{equation}
c_s= \frac{2g_2^4 + 3 g_Y^4/2}{\pi}S_T (-(2\alpha_2 +\alpha_Y)\sqrt{z}) +
 \frac{5g_2^4}{2\pi} S_T((\alpha_2-\alpha_Y)\sqrt{z})+
 \frac{6g_2^2 g_Y^2}{\pi} S_T (-(\alpha_2+\alpha_Y)\sqrt{z}),
 \label{eq:Sommerfeld}
\end{equation}
Using the modified $\gamma_A$ due to the corrections in $c_s$ as given in Eq.(\ref{eq:Sommerfeld}),  we solve unflavored Boltzmann equations (Eq.(\ref{be1}) - (\ref{be4}) ) as given in (\ref{app:rates})
and estimate the final lepton asymmetry.
\begin{figure*}[ht]
\includegraphics[width=0.49\textwidth,height=0.38\textwidth]{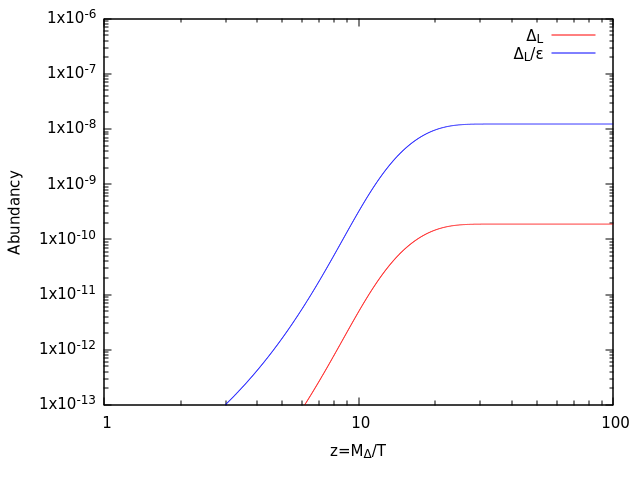}\hspace{0.158cm}
\includegraphics[width=0.49\textwidth,height=0.38\textwidth]{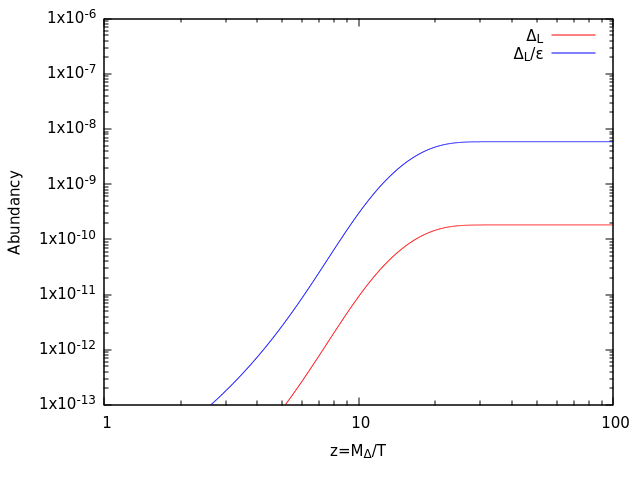} 
\vspace{-0.2cm}
 \caption{The figure shows the lepton asymmetry, $\Delta_{L}$ by solving Boltzmann equations for triplet mass, $M_{\Delta}=1.6$ TeV, and in the left (right) panel $B_{L}=0.95 (0.5)$. $\Delta_{L}$ is also plotted in units of $\epsilon$.}
 \label{fig:nep2}
\end{figure*}
We calculate the amount of CP violation using Eq.(\ref{eq:epsilon-resonant}) and it can be expressed as,
\begin{eqnarray}
\nonumber
 \epsilon_{\rm resonant}&=&\frac{(B_{L}B_{\phi})^{\frac{1}{2}}}{[a^{2}+2b^{2}+2c^{2}+d^{2}+2m^{2}+n^{2}]}\times\Biggl[a^{2}\sin(\phi_1-\alpha) +2b^{2}\sin(\phi_2-\alpha) 
 \\ 
 &+& 2c^{2}\sin(\phi_3-\alpha)+ d^{2}\sin(\phi_4-\alpha)  +2m^{2}\sin(\phi_5-\alpha) +n^{2}\sin(\phi_6-\alpha)\Biggr]
\end{eqnarray}
where, 
\begin{equation}
a^2 = \frac{A^2}{f(|\omega_{1}|,\omega_{2},\alpha,\phi_1 )}, ~
b^2 =\frac{B^2}{f(|\omega_{1}|,\omega_{2},\alpha,\phi_2 )}, ~
c^2 = \frac{C^2}{f(|\omega_{1}|,\omega_{2},\alpha,\phi_3 )},
\end{equation}

\begin{equation}
 d^2 =\frac{D^2}{f(|\omega_{1}|,\omega_{2},\alpha,\phi_4)},~
 m^2 = \frac{M^2}{f(|\omega_{1}|,\omega_{2},\alpha,\phi_5 )},~
 n^2= \frac{N^2}{f(|\omega_{1}|,\omega_{2},\alpha,\phi_6 )},~
\end{equation}

\begin{equation}
f(|\omega_{1}|,\omega_{2},\alpha,\phi_{i})=|\omega_{1}|^{2}+\omega^{2}_{2}+2|\omega_{1}|\omega_{2}\cos(\phi_{i}-\alpha).
\end{equation}
 We take the numerical values of $A, B, C, D, M,$ and $N$ from Eq.s (\ref{m1} - \ref{m6}) and from 
figures(\ref{fig:mdnplots}). The CP asymmetry  also depends on the phases $\phi_i$'s and $\alpha$. We scan over all the values of phases in ($0 - 2\pi$) range and choose those three sets of values that maximize the CP asymmetry parameter. The values are chosen in such a way that they also match with the numerical values of the corresponding branching ratio, $B_L$. We have chosen, $\omega_{1}=3$ GeV, $\omega_{2}=0.002$ GeV. We have taken the following set of values for calculating the CP asymmetry. 
\begin{itemize}
 \item Set I :  
 $\alpha=1.6, 
 \phi_{1}=\phi_{2}=1.5, 
 \phi_{3}=2.5, 
 \phi_{4}=\phi_{5}=5.5,
 \phi_{6}=2.5$.
 
 \item Set II : 
 $ \alpha=4.3,
 \phi_{1}=1.5,
 \phi_{2}=\phi_{3}=\phi_{4}=\phi_{5}=4.5,
 \phi_{6}=2.5$.
 
 \item Set III : 
 $ \alpha=4.4,
 \phi_{1}=3.5,
 \phi_{2}=1.5,
 \phi_{3}=4.5,
 \phi_{4}=2.5,
 \phi_{5}=\phi_{6}=5.5$.
\end{itemize}
 \begin{table}[h]
 \centering 
 \begin{tabular}{|p{1.5cm}||p{1cm}||p{2.5cm}||p{2cm}||p{2.5cm}|}
 \hline 
 Set&$B_{L}$&$\epsilon$&$\eta$&$\eta_{B}$\\
 \hline
  Set-I&$0.95$&$-1.536\times 10^{-2}$&$1.229\times 10^{-6}$&$5.474\times 10^{-10}$\\
  \hline 
  Set-II&$0.05$&$-1.299\times 10^{-2}$&$1.418\times 10^{-6}$&$5.342\times 10^{-10}$\\
 \hline
 Set-III&$0.5$&$-3.105\times10^{-2}$&$5.878\times 10^{-7}$&$5.293\times10^{-10}$\\
 \hline
\end{tabular}
\caption{Baryon asymmetries for different sets of $\epsilon$ and $B_{L}$, with $M_\Delta=1.6$ TeV.}
\label{table-TeV2}
\end{table}
For each of the choices of parameters we solve the set of Boltzmann equations
from Eq.s (\ref{be1} - \ref{be4}) and we plot the abundancy, $\Delta_L$  and find the efficiency
using Eq.(\ref{eq:effi}). The final baryon asymmetry is calculated using Eq.(\ref{eq:eta_b}).
Although we solve the Boltzmann equations for all three cases, in the Fig.(\ref{fig:nep2}) we have shown the results only for two 
cases: strongly hierarchical ($B_L =0.95$) and comparable branching ratios ($B_L =0.5$). 
However the results are summarised in the table (\ref{table-TeV2}) for all three cases. Here, one can notice that requiring baryon asymmetry to be in the
observational limit, the CP asymmetry parameter can be
maximized with certain choice of parameters. Also, it
can be seen, with hierarchical branching ratios the CP
asymmetry is suppressed and efficiency is enhanced 
as compared to the case of equal branching ratios. There is a significant suppression in the efficiency as compared to the the high scale leptogenesis scenario (as
discussed in section (\ref{sec:unflav})). This may be attributed to the Sommerfeld corrections in the rate of gauge scatterings, $\gamma_A$ as discussed in the beginning of this section. The quantitative results of the final baryon asymmetry, in this scenario, can further be improved by considering flavor effects which we shall explore in future work.
\section{Conclusion}
\label{sec:cncl}
We have studied baryogenesis through leptogenesis from the decay of scalar triplet in the SM, augmented with two triplet scalars. 
The coexistence of two decay processes of lighter triplet and hence two branching ratios
($B_L$ and $B_\phi$) introduce new features in leptogenesis as compared to
other scenarios. The amount of baryon asymmetry generation depends on the strength of
CP violation and efficiency of the CP violating decay. The CP asymmetry parameter is determined by the complex phases of the Yukawa matrix and vev of the triplet scalar through the neutrino mass matrix, triplet mass, and branching ratios of triplet scalars to the doublet scalars and the SM leptons. It is known that strongly hierarchical branching ratios enhance the efficiency but suppress the CP violation.
The generation of baryon asymmetry then requires the mass of the triplets to be between
$\sim 10^9 - 10^{12}$ GeV. In our study, we find that for a sizable amount of CP asymmetry
($\sim 10^{-7}$) and efficiency ($\sim 10^{-2}$) the triplet mass can be as low as $1\times 10^{10}$ GeV. In this case, the order of magnitude of generated baryon asymmetry matches with 
the observed value. The reported value of triplet mass is suitable for studying the flavor
effects in leptogenesis. 
The CP asymmetries arise  because of the interference of the tree level diagram with  one loop diagrams mediated by doublet Higgs and leptons.
The former one is both lepton number violating as well as lepton flavor violating ($\epsilon^{L_i (\slashed{L}, \slashed{F})}$) and the latter one is only lepton flavor violating ($\epsilon^{L_i(\slashed{F})}$).
The purely flavored, $\epsilon^{L_i(\slashed{F})}$ dominates over the $\epsilon^{L_i (\slashed{L}, \slashed{F})}$ in the case where triplets couple to leptons more than the doublet Higgs.
This ensures the hierarchical branching ratios, $B_L \gg B_\phi$. In our study, the contributions to CP asymmetry parameters are determined from the elements of the Yukawa coupling matrix which in 
turn are determined from neutrino mass matrix elements. In our study, the CP asymmetry arising from the purely lepton flavor violating part comes out to be negligible as compared to the lepton flavor and number violating part. So,
producing adequate BAU seems not attainable for purely flavored leptogenesis. However, the other choice of $B_\phi \gg B_L$ and flavor effects show to meet the objective with the proper choice of 
CP violating phases of the vev of the triplet Higgs scalar and other phases of neutrino mass matrix. The phase of the complex vev of the triplet scalar is
constrained from analysis from scalar potential as $\alpha \neq n\pi, n=  1,2, 3,\cdots$. Also, requiring non-zero CP violation, the phase $\alpha$ is 
constrained.
We also study the possibility of TeV scale leptogenesis taking resonant effects into account where the two triplet scalars can be nearly degenerate in their masses.

Triplet scalars arise naturally in many models beyond the SM such as left-right symmetric model. They offer rich phenomenology in energy frontiers like the Large Hadron Collider and
processes like neutrinoless double beta decay and lepton flavor violation.  
They have
also been studied combined with inflation \cite{Barrie:2021mwi} and strong CP violation \cite{Bertolini:2014aia}. Adding triplets to the SM
and augmenting with different symmetries explain the origin of different textures in the neutrino mass matrix.  
The high scale leptogenesis mechanisms with triplet scalars  can now be tested with Cosmological 
colliders physics \cite{Cui:2021iie,Lu:2019tjj} in the light of CMB and Large Scale Structure observations targeting primordial non-Gaussianity. It can also be tested through gravitational waves \cite{Dror:2019syi}. It offers to study combinations of
above-laid avenues in future.
\newpage

\appendix
\section{Neutrino mass matrix elements}
\label{sec:elements}
Using Eq.(\ref{eq:conjecture}) in Eq.(\ref{eq:mneu1}), the elements of neutrino
mass matrix are given by,
\begin{equation}\nonumber
 A^{2}=a^{2}[|\omega_{1}|^{2}+\omega^{2}_{2}+2|\omega_{1}|\omega_{2}\cos(\phi_{a}-\phi_{a'}+\alpha)],
 \label{aA}
\end{equation}
\begin{equation}\nonumber
 \phi_{A}={\rm tan}^{-1}\left[\frac{|\omega_{1}| \sin(\phi_{a}+\alpha)+\omega_{2}\sin\phi_{a'}}{|\omega_{1}|\cos(\phi_{a}+\alpha)+\omega_{2}\cos\phi_{a'}}\right],
\end{equation}
\begin{equation}\nonumber
 B^{2}=b^{2}[|\omega_{1}|^{2}+\omega^{2}_{2}+2|\omega_{1}|\omega_{2} \cos(\phi_{b}-\phi_{b'}+\alpha)],
 \label{bB}
\end{equation}
\begin{equation}\nonumber
 \phi_{B}={\rm tan}^{-1}\left[\frac{|\omega_{1}|\sin(\phi_{b}+\alpha)+\omega_{2}\sin\phi_{b'}}{|\omega_{1}|\cos(\phi_{b}+\alpha)+\omega_{2}\cos\phi_{b'}}\right],
\end{equation}
\begin{equation}\nonumber
 C^{2}=c^{2}[|\omega_{1}|^{2}+\omega^{2}_{2}+2|\omega_{1}|\omega_{2} \cos(\phi_{c}-\phi_{c'}+\alpha)],
 \label{cC}
\end{equation}
\begin{equation}\nonumber
 \phi_{C}={\rm tan}^{-1}\left[\frac{|\omega_{1}|\sin(\phi_{c}+\alpha)+\omega_{2}\sin\phi_{c'}}{|\omega_{1}|\cos(\phi_{c}+\alpha)+\omega_{2}\cos\phi_{c'}}\right],
\end{equation}
\begin{equation}\nonumber
 D^{2}=d^{2}[|\omega_{1}|^{2}+\omega^{2}_{2}+2|\omega_{1}|\omega_{2} \cos(\phi_{d}-\phi_{d'}+\alpha)],
 \label{dD}
\end{equation}
\begin{equation}\nonumber
 \phi_{D}={\rm tan}^{-1}\left[\frac{|\omega_{1}|\sin(\phi_{d}+\alpha)+\omega_{2}\sin\phi_{d'}}{|\omega_{1}|\cos(\phi_{d}+\alpha)+\omega_{2}\cos\phi_{d'}}\right],
\end{equation}
\begin{equation}\nonumber
 M^{2}=m^{2}[|\omega_{1}|^{2}+\omega^{2}_{2}+2|\omega_{1}|\omega_{2} \cos(\phi_{m}-\phi_{m'}+\alpha)],
 \label{mM}
\end{equation}
\begin{equation}\nonumber
 \phi_{M}={\rm tan}^{-1}\left[\frac{|\omega_{1}|\sin(\phi_{m}+\alpha)+\omega_{2}\sin\phi_{m'}}{|\omega_{1}|\cos(\phi_{m}+\alpha)+\omega_{2}\cos\phi_{m'}}\right],
\end{equation}
\begin{equation}\nonumber
 N^{2}=n^{2}[|\omega_{1}|^{2}+\omega^{2}_{2}+2|\omega_{1}|\omega_{2} \cos(\phi_{n}-\phi_{n'}+\alpha)],
 \label{nN}
\end{equation}
\begin{equation}
 \phi_{N}={\rm tan}^{-1}\left[\frac{|\omega_{1}|\sin(\phi_{n}+\alpha)+\omega_{2}\sin\phi_{n'}}{|\omega_{1}|\cos(\phi_{n}+\alpha)+\omega_{2}\cos\phi_{n'}}\right].
 \label{eq:A-to-N}
\end{equation}
For unflavoured Leptogenesis, we choose benchmark points related to $B_{L}=0.9995$ that correspond to  $\epsilon=1.5\times 10^{-7}$. The corresponding phase values used are 
 $\phi_{a}=1.20$, $\phi_{b}=1.00$, $\phi_{c}=1.30$, $\phi_{d}=1.10$, $\phi_{m}=0.30$, $\phi_{n}=1.50$ and  
 $\phi_{a'}=5.31999969$, $\phi_{b'}=5.11999941$, $\phi_{c'}=5.41999912$, $\phi_{d'}=5.21999931$, $\phi_{m'}=4.41999960$, $\phi_{n'}=5.61999941$ so that  
 $\phi_{a'}-\phi_{a}=\phi_{b'}-\phi_{b}=\phi_{c'}-\phi_{c}=\phi_{d'}-\phi_{d}=\phi_{m'}-\phi_{m}=\phi_{n'}-\phi_{n}=\phi=4.11999941$ and for $\alpha=2.8499976$,
 the Yukawa matrices can be calculated as  
 \begin{equation}
  Y^{(1)}_{\Delta}=10^{-12}\times
  \begin{pmatrix}
   3.42+i8.79 & 2.25+i3.50 & 1.40+i5.04\\
   2.25+i3.50 & 2.84+i5.57 & 5.70+i1.76\\
   1.40+i5.04 & 5.70+i1.76 & 0.38+i5.42
  \end{pmatrix},
 \end{equation}

 \begin{equation}
  Y^{(2)}_{\Delta}=10^{-12}\times
  \begin{pmatrix}
   5.38-i7.74 & 1.65-i3.82 & 3.40-i3.97\\
   1.65-i3.82 & 3.04-i5.46 & -1.72-i5.72\\
   3.40-i3.97 & -1.72-i5.72 & 4.28-i3.34
  \end{pmatrix}.
 \end{equation}
 Similarly for flavored leptogenesis, for the benchmark points
 $B_{\phi}=0.9995$ that gives rise to flavored CP asymmetries,
 $\epsilon^{a}=-1.834\times 10^{-8}$, $\epsilon^{\tau}=-3.293\times 10^{-9}$ and corresponds to the phase values
  $\phi_{a}=1.00$, $\phi_{b}=2.00$, $\phi_{c}=1.20000005$, $\phi_{d}=1.10000002$, $\phi_{m}=0.300000012$, $\phi_{n}=1.10000002$ and  
 $\phi_{a'}=2.5$, $\phi_{b'}=3.5$, $\phi_{c'}=2.70000005$, $\phi_{d'}=2.59999990$, $\phi_{m'}=2.00000024$, $\phi_{n'}=6.19999743$ and  
 $\phi_{a'}-\phi_{a}=\phi_{1}=1.50$, $\phi_{b'}-\phi_{b}=\phi_{2}=1.50$, $\phi_{c'}-\phi_{c}=\phi_{3}=1.50$, $\phi_{d'}-\phi_{d}=\phi_{4}=1.50$, $\phi_{m'}-\phi_{m}=\phi_{5}=1.70000029$, $\phi_{n'}-\phi_{n}=\phi_{6}=5.09999752$ for 
 $\alpha=2.49999976$ the Yukawa matrices are calculated to be, 
 \begin{equation}
  Y^{(1)}_{\Delta}=10^{-12}\times
  \begin{pmatrix}
   5.10+i7.94 & -1.73+i3.79 & 1.89+i4.87\\
   -1.73+i3.79 & 2.84+i5.57 & 5.70+i1.76\\
   1.89+i4.87 & 5.70+i1.76 & 2.47+i4.85
  \end{pmatrix},
 \end{equation}

 \begin{equation}
  Y^{(2)}_{\Delta}=10^{-12}\times
  \begin{pmatrix}
   -7.56+i5.64 & -3.90-i1.46 & -4.73+i2.23\\
   -3.90-i1.46 & -5.36+i3.22 & -2.48+i5.43\\
   -4.73+i2.23 & -2.48+i5.43 & 5.42-i0.45
  \end{pmatrix}.
 \end{equation}
 It is clear from the study that even though the two Yukawa coupling matrices share common moduli of the elements, the complex elements are different due to the associated phases. 
\section{Reaction Densities}
\label{app:rates}
\begin{figure}[h]
 \centering
\includegraphics[scale=0.5]{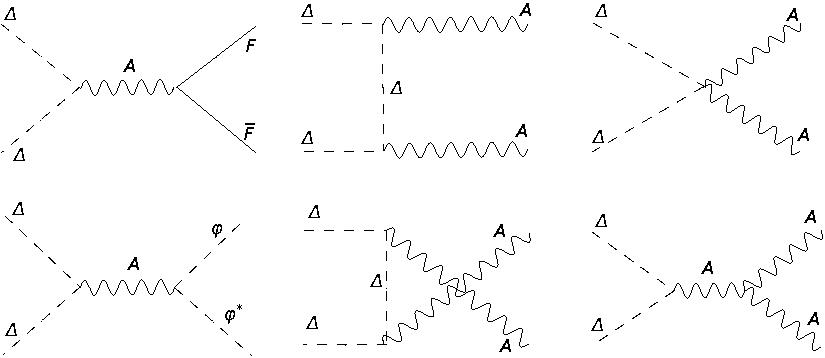}
\caption{Feynman diagrams that contribute to the interaction rate $\gamma_A$}
\label{gammaA}
\end{figure}

  The  space-time density of total decay of triplet scalar $\Delta$ and its anti-particle $\overline{\Delta}$, $\gamma_D$ and  that of $2\leftrightarrow2$ scattering, $\gamma_A$ are  given by,
 \begin{equation}
 \gamma_D=s\Gamma_\Delta\Sigma^{eq}_\Delta\frac{K_1(z)}{K_2(z)},
 \end{equation}
 \begin{equation}
 \gamma_A =\frac{M^4_\Delta}{64\pi^4}\int^{\infty}_{x_{\rm min}} dx  \sqrt{x}\frac{K_1\left( z \sqrt{x} \right) \hat{\sigma}_S}{z}.
 \label{eq:gamma-gen}
 \end{equation}
 Here, $x= {\rm s}/M^2_\Delta$, $\textbf{s}$ is the Mandelstam variable, square of center of mass energy \cite{Halzen:1984mc}, $\hat{\sigma}_S$ is the reduced cross section of $2\rightarrow 2$ scatterings (relevant diagrams in fig.(\ref{gammaA})). For processes mediated by gauge bosons $x_{\rm min} =4$ and for Yukawa induces reactions $x_{\rm min} =0$. For processes mediated by gauge bosons the reduced cross section, $\hat{\sigma}_S$ is given by Eq. (\ref{eq:sigma-S}) and the corresponding rate density of scattering is given by,
 \begin{equation}
 \gamma_A=\frac{M_\Delta T^3e^{-\frac{2M_\Delta}{T}}}{64\pi^4}(9g^4+12g_2^2g^2_{\mathcal{Y}}+3g^4_Y)\left(1+\frac{3T}{4M_\Delta}\right)
 \label{eq:gammaA}
 \end{equation}
Here we take $\Gamma_{\Delta}=\frac{M_\Delta^2 \tilde{m}_\Delta}{16\pi v^2 \sqrt{B_L B_{\phi}}}$, $\tilde{m}_\Delta = {\rm Tr} (m^{(2)\dagger}_\nu m^{(2)}_\nu )$, $\Sigma^{\rm eq}_\Delta=2Y^{\rm eq}_\Delta=2\times\frac{45g_\Delta}{4\pi^4g_{*}}z^2K_2(z)$ and
$g_\Delta=3$, $g_{*}=106.75$ are the triplet scalar and SM particle-degrees of freedom, respectively \cite{Felipe:2013kk}. 
$g_2, g_{\mathcal{Y}}$ are the coupling constants corresponding to $SU(2)_L$ and $U(1)_{\mathcal{Y}}$ interactions, respectively, having values, $g_2 =\frac{e}{\sin\theta_W}=0.651742$ and $g_{\mathcal{Y}}=\frac{e}{\cos\theta_W}=0.461388$.  The other reaction densities, $\gamma$'s are to be calculated using the integration given in
Eq. (\ref{eq:gamma-gen}).

 The unflavored Boltzmann equations are given by \cite{Hallgren:2007nq, Hambye:2005tk}, 
 \begin{equation}
 sHz\frac{d\Sigma_\Delta}{dz}=-\left(\frac{\Sigma_\Delta}{\Sigma^{eq}_\Delta}-1\right)\gamma_D-2\left(\frac{\Sigma^2_\Delta}{\Sigma^{2eq}_\Delta}-1\right)\gamma_A,
 \label{be1}
 \end{equation}
 \begin{equation}
 sHz\frac{d\Delta_L}{dz}=\gamma_D\epsilon\left(\frac{\Sigma_\Delta}{\Sigma^{eq}_\Delta}-1\right)-2\gamma_DB_L \left(\frac{\Delta_L}{Y^{eq}_L}+\frac{\Delta_\Delta}{\Sigma^{eq}_\Delta}\right),
 \label{be2}
 \end{equation}
 \begin{equation}
 sHz\frac{d\Delta_\phi}{dz}=\gamma_D\epsilon\left(\frac{\Sigma_\Delta}{\Sigma^{eq}_\Delta}-1\right)-2\gamma_DB_
 {\phi}\left(\frac{\Delta_{\phi}}{Y^{eq}_{\phi}}-\frac{\Delta_\Delta}{\Sigma^{eq}_\Delta}\right),
 \label{be3}
 \end{equation}
 \begin{equation}
 sHz\frac{d\Delta_\Delta}{dz}=-\gamma_D\left(\frac{\Delta_\Delta}{\Sigma^{eq}_\Delta}+B_L\frac{\Delta_L}{Y^{eq}_L}-B_{\phi}\frac{\Delta_{\phi}}{Y^{eq}_{\phi}}\right).
 \label{be4}
 \end{equation}
The Boltzmann equations for triplet leptogenesis has certain 
distinct features: (1) scalar triplet is not a self-conjugate state, unlike heavy Majorana neutrinos. So a non-vanishing asymmetry arises between the triplet ($\Delta$) and anti-triplet ($\overline{\Delta}$) abundances. It definitely affects the leptogenesis dynamics differently. That is why, one extra Boltzmann equation, corresponding to $\Delta_\Delta$ (Eq.s (\ref{be1})-(\ref{be4})), appears in this scenario.
(2) The triplet scalar is charged under $SU(2)_L\times U(1)_{\mathcal{Y}}$. So the triplet and anti-triplet are able to annihilate before they take part in decay. A significantly large lepton asymmetry can be generated only if the decay rates are higher than the rate of gauge annihilation, which are typically in thermal equilibrium. But, it seems to contradict Sakharov's third condition: the condition of being out of thermal equilibrium. But as triplet has several decay channels, it is possible for some of those to be out of equilibrium and that resolves the mentioned conflict. Again, the final lepton asymmetry does not depend on the initial conditions, as the gauge scatterings keep the triplet scalar abundance $\Sigma_\Delta$ close to thermal equilibrium. Thereby, it can be observed that the Boltzmann equation corresponding to $\Sigma_\Delta$ is dominated by $\gamma_D$, which puts $\Sigma_\Delta$ close to $\Sigma^{eq}_\Delta$, whereas gauge scatterings have a negligible effect on it.
 \section{Sommerfeld Corrections}
 \label{app:sommer}
 For a single abelian mass-less vector with potential $V=\alpha/r$, the Sommerfeld
correction to the reduced cross section, $\sigma= S(x) \sigma_{\rm perturbative}$ where, 
\begin{equation}
 S(x) = \frac{-\pi x}{1-e^{\pi x}},\quad x= \alpha/\beta,
\end{equation}
where, $\beta =\sqrt{1-4M^2_\Delta}$ is the triplet velocity
in $\Delta \Delta^*$ center of mass frame, $\alpha < 0 (>0)$
describes the attractive (repulsive) potential.
The thermally averaged value of Sommerfeld correction is,
\begin{equation}
 S_T(y) = \frac{\int_0 \beta^2 e^{-M_{\Delta}\beta^2/T} S(\alpha/\beta)d\beta}{\int_0 \beta^2 e^{-M_{\Delta}\beta^2/T}d \beta}, ~ y\equiv \alpha/ \beta_T,
\end{equation}
where, $\beta_T\equiv \sqrt{T/M_{\Delta}}$ is characteristic velocity of the thermal bath. The 
function $S_T$ can be computed numerically and leads to $S_T(x) \approx S(x)$. 
The generalization to massive vectors needs a list of potential and annihilation
matrices and has been discussed in Ref. \cite{Hisano:2006nn}. For $SU(2)_L$ invariant
limit ($M \gg M_{W,Z}$), the Sommerfeld correction was calculated in ref. \cite{Strumia:2008cf}.
The isospin of two-body triplet states $\Delta\Delta^*$ have total isospin $I={1,3,5}$, as it can be decomposed as $3\bigotimes 3 = 1_{S} \bigoplus 3_A \bigoplus 5_S$ which can result from total spin $s=0, 1$ and total angular momentum $L$. The potentials are, 
$V= - 2\alpha_2/r, -\alpha_2/r, \alpha_2/r$ for singlet, triplet and quintuplet states respectively. Since in the energy regime we are interested, $s$-wave $(L=0)$ annihilations are important, the final two SM states can have isospin $I={1,3,5}$. The correction to 
annihilation rate from Sommerfeld corrections is given in Eq.(\ref{eq:Sommerfeld}), where the first, second and third term on right hand side of the equation correspond
to final isospin state $I=1,2$ and $3$ respectively. In the limit $S_T \rightarrow 1$,
we get back the annihilation rate for given in Eq. (\ref{eq:gammaA}) in the (\ref{app:rates}). 
\begin{figure}[h]
  \centering
  \begin{subfigure}{0.45\linewidth}
   \includegraphics[width=\linewidth]{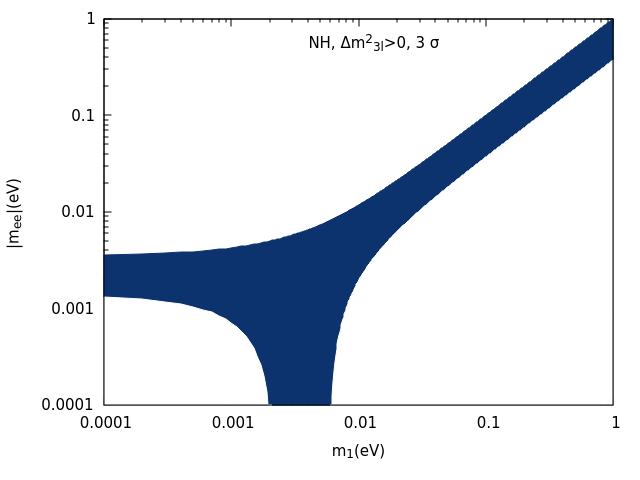}
   \caption{$m_{ee}$ vs $m_1$}
   \label{fig:st23d}
   \end{subfigure}
   \begin{subfigure}{0.45\linewidth}
   \includegraphics[width=\linewidth]{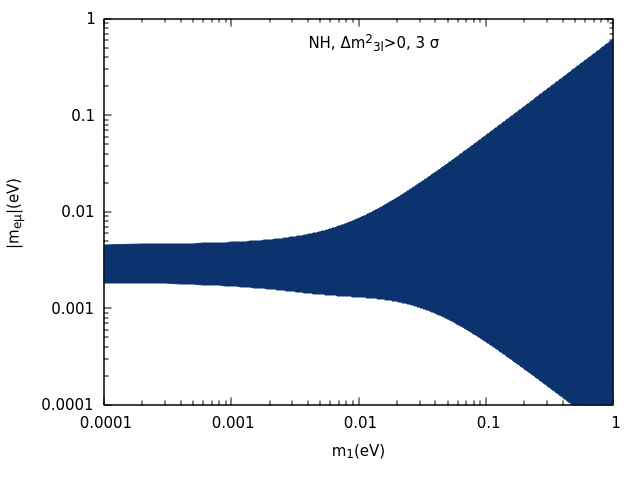}
   \caption{$m_{e\mu}$ vs $m_1$}
   \label{fig:smee}
   \end{subfigure}
   \begin{subfigure}{0.45\linewidth}
   \includegraphics[width=\linewidth]{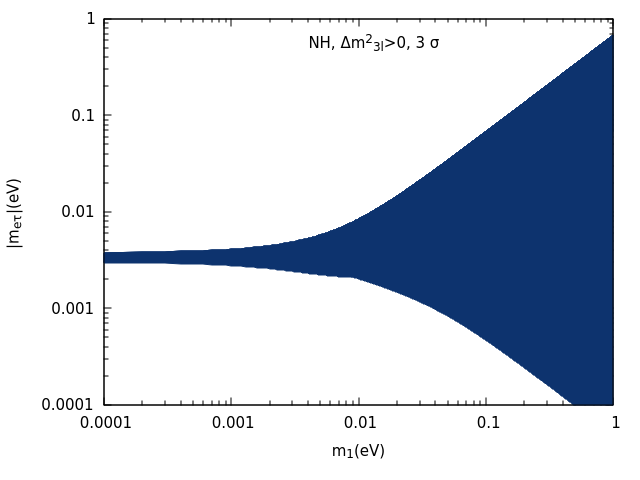}
   \caption{$m_{e\tau}$ vs $m_1$}
   \label{fig:m1mee}
   \end{subfigure}
   \begin{subfigure}{0.45\linewidth}
   \includegraphics[width=\linewidth]{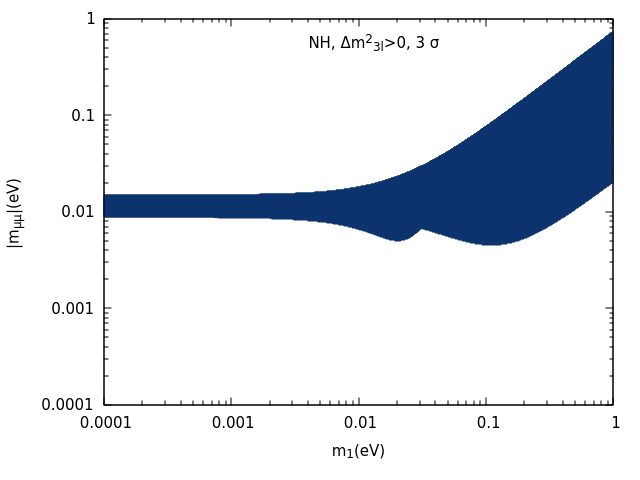}
   \caption{$m_{\mu\mu}$ vs $m_1$}
   \label{fig:m1s}
   \end{subfigure}
   \begin{subfigure}{0.45\linewidth}
   \includegraphics[width=\linewidth]{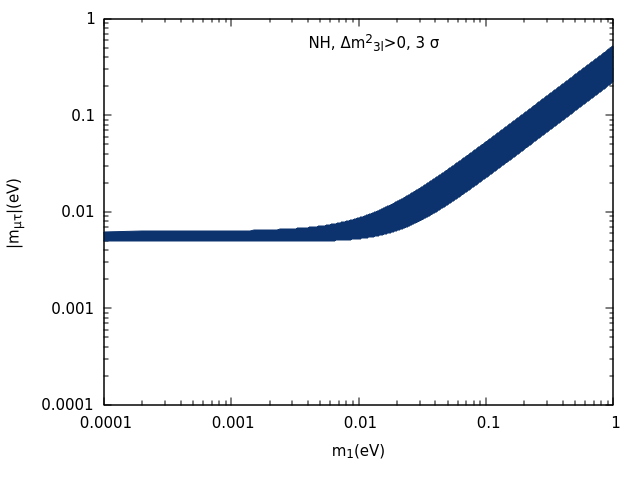}
   \caption{$m_{\mu\tau}$ vs $m_1$}
   \label{fig:mdnrr}
   \end{subfigure}
   \begin{subfigure}{0.45\linewidth}
   \includegraphics[width=\linewidth]{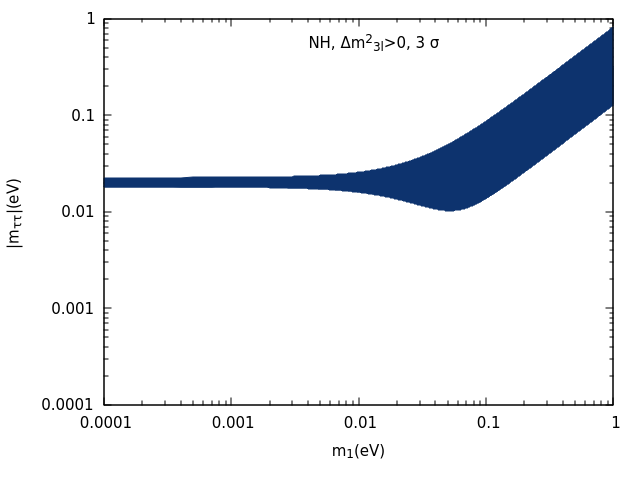}
   \caption{$m_{\tau\tau}$ vs $m_1$}
   \label{fig:mdnsr}
   \end{subfigure}
    \caption{Correlation plots for elements of neutrino mass matrix.}
   \label{fig:mdnplots}
   \end{figure}

\begin{figure}[ht]
 \centering
     \begin{subfigure}{0.4\linewidth}
   \includegraphics[width=\linewidth]{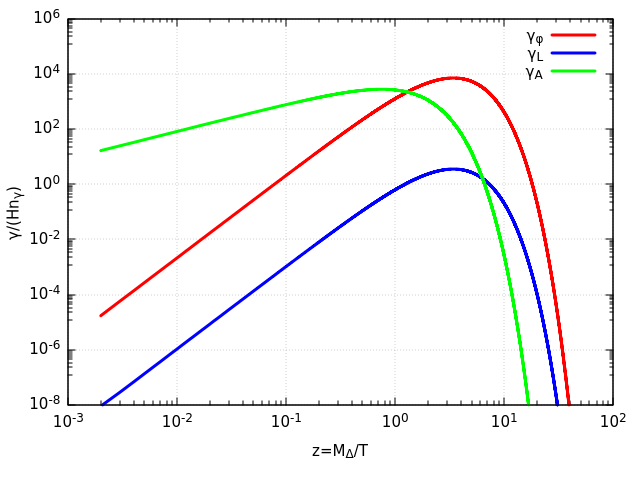}
   \caption{$B_{L}=0.0005,B_{\phi}=0.9995$}
   \label{fig:lmt2}
  \end{subfigure}
  \begin{subfigure}{0.4\linewidth}
   \includegraphics[width=\linewidth]{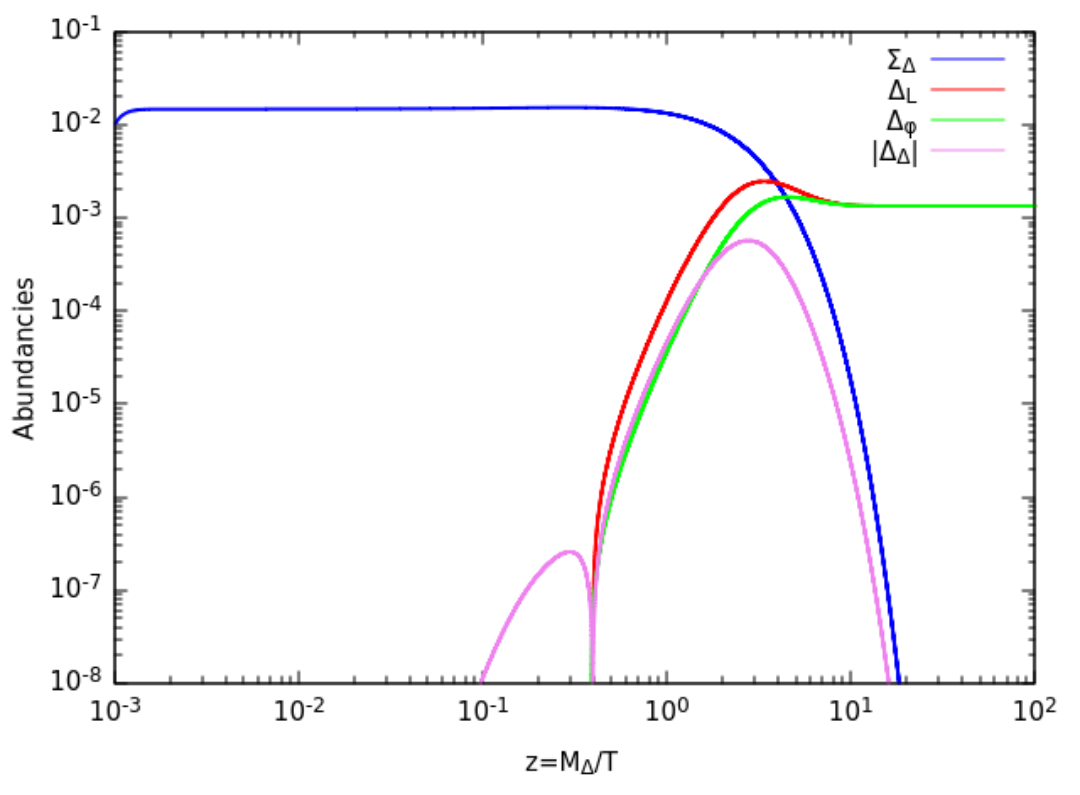}
   \caption{$B_{L}=0.0005,B_{\phi}=0.9995$}
   \label{fig:blbh12}
  \end{subfigure}
  \begin{subfigure}{0.4\linewidth}
   \includegraphics[width=\linewidth]{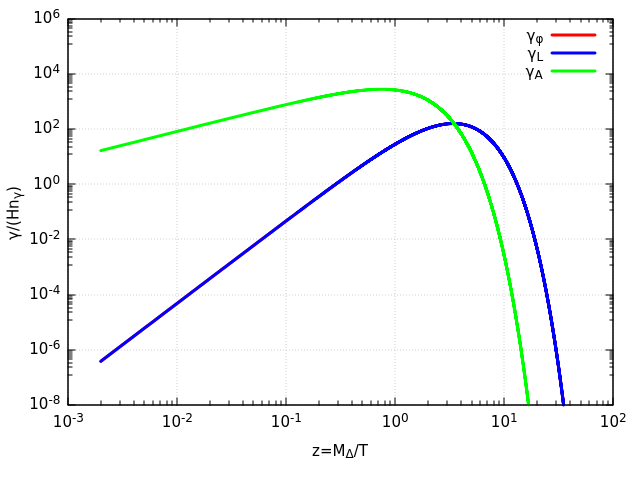}
   \caption{$B_{L}=0.5000,B_{\phi}=0.5000$}
   \label{fig:lmt5}
  \end{subfigure}
  \begin{subfigure}{0.4\linewidth}
   \includegraphics[width=\linewidth]{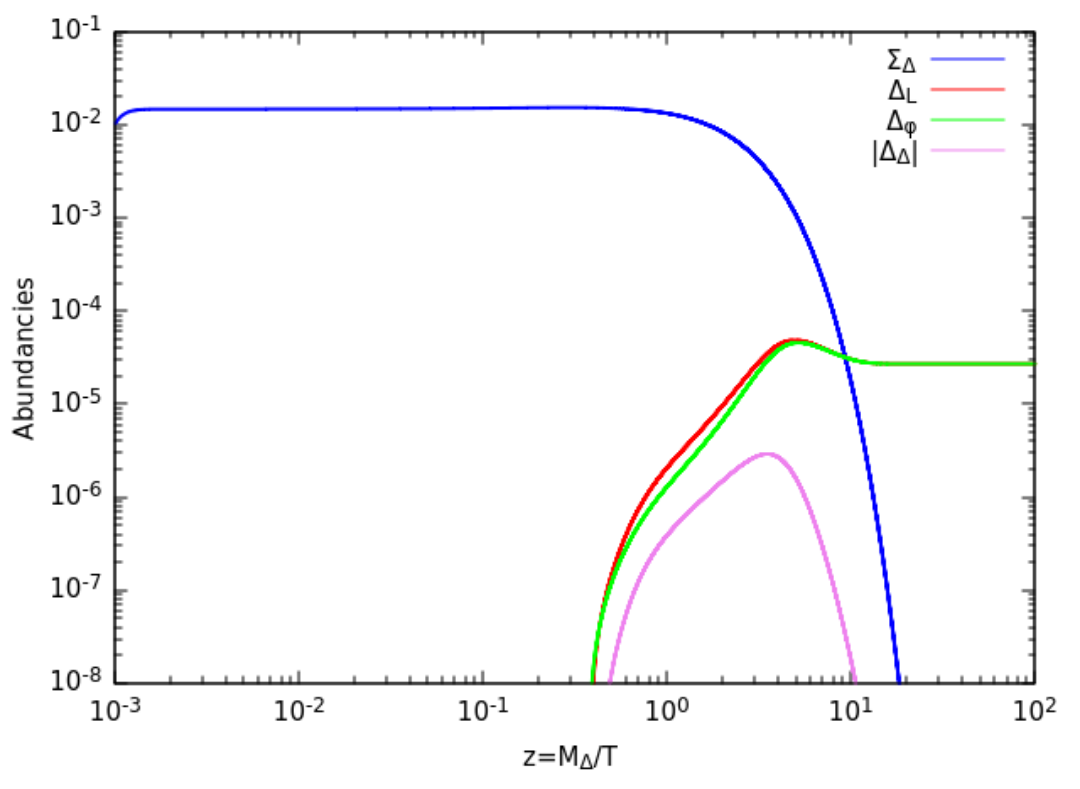}
   \caption{$B_{L}=0.5000,B_{\phi}=0.5000$}
   \label{fig:blbh2}
  \end{subfigure}
    \medskip
  \small
  \caption{Dependence of decay rates on $B_{L}$ and $B_{\phi}$ and corresponding abundancy plots.
    On the left column, the dependence of decay rates on $B_{L}$ and $B_{\phi}$ is studied; on the right column, corresponding dependence of abundancies on $B_{L}$ and $B_{\phi}$ is studied for $M_{\Delta}=5\times10^{10}$ GeV and 
    $\tilde{m}_{\Delta}=0.05$ eV. The abundancies ($\Delta_{L}$, $\Delta_{\phi}$, $|\Delta_{\Delta}|$) are plotted in units of $\epsilon$.} 
\label{fig:blbp}
 \end{figure}

\end{document}